 \documentclass[reprint,amsmath,amssymb,aps,pra]{revtex4-2}

\usepackage{amsmath,amssymb,amsfonts,amstext,graphics,graphicx,subfigure,dcolumn,bm}
\usepackage{setspace}
\usepackage{mathtools}
\usepackage[colorlinks]{hyperref}
\usepackage{color}

\usepackage{mathbbol}
\usepackage{natbib}

\usepackage{enumerate}

\newcommand{\KN}{\mathbin{\bigcirc\mspace{-15mu}\wedge\mspace{3mu}}}

\newtheorem{theorem}{Theorem}

\newtheorem{definition}[theorem]{Definition}

\newtheorem{lemma}[theorem]{Lemma}

\newtheorem{proposition}[theorem]{Proposition}
\newtheorem{remark}[theorem]{Remark}

\newenvironment{proof}[1][Proof]{\noindent\textbf{#1.} }{\ \rule{0.5em}{0.5em}}

\def\XXint#1#2#3{{\setbox0=\hbox{$#1{#2#3}{\int}$ }
\vcenter{\hbox{$#2#3$ }}\kern-.5\wd0}}

\def\cala{\mathcal{A}}

\def\calh{\mathcal{H}}

\def\cals{\mathcal{S}}

\def\C{\mathbb{C}}

\def\R{\mathbb{R}}
\def\C{\mathbb{C}}

\def\Dt#1{\frac{\mathrm{d} #1}{\mathrm{d}t}}

\def\bq{\begin{equation}}
\def\eq{\end{equation}}
\def\bqy{\begin{eqnarray}}
\def\eqy{\end{eqnarray}}

\def\bal#1\eal{\begin{align}#1\end{align}}

\def\be{\beta}

\def\de{\delta}

\def\ga{\gamma}

\def\la{\lambda}

\def\na{\nabla}

\def\Om{\Omega}

\def\si{\sigma}
\def\Si{\Sigma}

\def\bfB{\mathbf{B}}

\def\bfA{\mathbf{A}}

\def\bfM{\mathbf{M}}
\def\bfH{\mathbf{H}}

\def\p{\partial}
\def\na{\nabla}

\def \Tr {\operatorname{Tr}}

\def\nn{\nonumber}

\def\bq{\begin{equation}}
\def\eq{\end{equation}}
\def\bqy{\begin{eqnarray}}
\def\eqy{\end{eqnarray}}

\def\bal#1\eal{\begin{align}#1\end{align}}

\def\4tensLa{\bar{\bar{\Lambda}}}

\def\be{\beta}

\def\de{\delta}

\def\ga{\gamma}

\def\la{\lambda}

\def\na{\nabla}

\def\Om{\Omega}

\def\si{\sigma}
\def\Si{\Sigma}

\def\md{\mathrm{d}}

\def\p{\partial}
\def\na{\nabla}

\def\nn{\nonumber}

\begin{document}
 
\title{Thermodynamically Consistent Quantum–Classical Hybrid Dynamics: A Metriplectic Approach to Generalized Lindblad and Bloch–Torrey Equations}
\author{Azeddine Zaidni}
\affiliation{Mohammed VI Polytechnic University, College of Computing, Lot 660, Hay Moulay Rachid, Ben Guerir, 43150, Morocco}
\email{AZEDDINE.ZAIDNI@um6p.ma}
\author{Ayoub Arraji}
\email{Ayoub.Arraji@um6p.ma}
\affiliation{Mohammed VI Polytechnic University, College of Computing, Lot 660, Hay Moulay Rachid, Ben Guerir, 43150, Morocco}
\author{Philip J.~Morrison}
\email{morrison@physics.utexas.edu}
\affiliation{Department of Physics and Institute for Fusion Studies, 
The University of Texas at Austin, Austin, TX, 78712, USA}

\date{\today}

\begin{abstract}

\bigskip

We develop a  metriplectic framework for quantum--classical hybrid dynamics that ensures thermodynamic consistency through conservation of energy and non-negative entropy production. The approach combines a hybrid Poisson bracket with a metriplectic $4$-bracket, with coupling between quantum and classical subsystems arising solely through dissipative processes. Within this hybrid formulation, we produce thermodynamically consistent  generalizations of the Lindblad master equations and the Bloch--Torrey equations for open quantum systems. This provides a unified description of quantum relaxation and classical diffusion, together with a clear interpretation of entropy production.

\end{abstract}

\maketitle

\begin{widetext}

\tableofcontents

\end{widetext}


\section{Introduction}

The consistent description of systems involving both classical and quantum degrees of freedom remains a longstanding challenge in theoretical physics. Such hybrid systems naturally arise in a wide variety of contexts, including semiclassical gravity, where the gravitational field is sourced by the expectation value of the quantum energy--momentum tensor, collective motion in electromagnetic fields, quantum measurement theory, and the dynamics of quantum systems interacting with macroscopic environments. In all these situations, one is confronted with the problem of coupling fundamentally different dynamical structures: classical Hamiltonian mechanics and quantum unitary evolution.

In the canonical formulation of quantum mechanics, the time evolution of an isolated system is governed by unitary dynamics, implying conservation of energy and preservation of phase coherence. However, realistic physical systems are never perfectly isolated and inevitably interact with their surroundings. Such interactions lead to dissipation, decoherence, and entropy production, which are essential for understanding phenomena such as spontaneous emission, thermalization, and the operation of quantum technological devices \cite{breuer2002theory,colombo2025}. These effects are typically modeled by considering the quantum subsystem as open and embedding it into an external environment, often treated in an effective or phenomenological manner.

The formulation of a unified dynamical framework for systems combining classical and quantum degrees of freedom is therefore of fundamental importance. Early attempts aimed at constructing Hamiltonian descriptions of hybrid systems \cite{antoniou1999hamiltonian}, and a variety of approaches have since been developed, including mean-field models, Ehrenfest dynamics, and quantum--classical Liouville equations. In many of these formulations, the total Hamiltonian is taken as the sum of classical and quantum contributions, and the dynamics is generated by a combination of classical and quantum Poisson brackets \cite{elze2012linear,caro1999impediments,peres2001hybrid,buric2012statistical}. Other approaches incorporate coupling at the nondissipative (Hamiltonian) level, often assuming additive structures for energy and entropy \cite{alonso2020entropy}, or more general non-additive constructions involving both classical and quantum densities as dynamical variables \cite{tronci2023lagrangian,tronci2025entropy}.

Despite these efforts, there is no universally accepted definition of a quantum--classical hybrid system, and many existing models are best understood as approximations to an underlying fully quantum description. In particular, ensuring thermodynamic consistency---namely, the simultaneous conservation of total energy and monotonic increase of entropy---remains a delicate issue. Moreover, most approaches focus primarily on the nondissipative aspects of the dynamics, while the role of dissipation in mediating the interaction between classical and quantum subsystems has received comparatively less attention.

In this paper we will use the term nondissipative to refer to the Hamiltonian part of the dynamics, which is often referred to as reversible.  We will use dissipative to refer to the entropy producing part of the dynamics, which is often referred to as irreversible.

In parallel, the development of the \emph{metriplectic  formalism} \cite{pjm84,pjm84b,pjm86,pjmU24} (also called 
\scalebox{.7}{GENERIC} in \cite{og97}) has provided a powerful framework for constructing thermodynamically consistent dynamical systems. The metriplectic approach extends Hamiltonian dynamics by incorporating an additional symmetric bracket structure that generates dissipation while preserving fundamental invariants. Since its introduction in the 1980s, metripletic dynamics has more recently been successfully applied to a wide range of classical systems, from finite-dimensional models such as rigid bodies and oscillators to infinite-dimensional systems such as fluids and plasmas (see, e.g., \cite{pjmM18,pjmC20,pjmU24,pjmZB24,pjmS24,pjmBZ25,pjmZ25,pjmS25}). (See \cite{pjmU24}  for discussion and comparison to some diﬀerent approaches to nonequilibrium thermodynamics, e.g., \cite{pjmK82,grmela1984particle,kaufman1984dissipative}.) Also, extensions to quantum systems have been explored, where dissipation is typically modeled through Lindblad-type operators  \cite{lindblad1976generators,gorini1976completely} (see, e.g., \cite{ottinger2010nonlinear,ottinger2011geometry} for such an exploration).

In contrast to most existing approaches, we focus on a regime in which the coupling between the classical and quantum subsystems is mediated exclusively through the {dissipative} part of the dynamics, while nondissipative interactions are neglected. This regime is relevant for strongly dissipative environments where decoherence suppresses coherent back-reaction between classical and quantum subsystems.

Within this framework, the nondissipative dynamics is generated by a hybrid Poisson bracket constructed as the direct sum of classical and quantum contributions, while the dissipative  dynamics is described by a hybrid metriplectic $4$-bracket \cite{pjmU24}. This construction ensures that the total energy of the combined system is conserved, while the total entropy is non-decreasing, thereby satisfying the first and second laws of thermodynamics at the hybrid level. The resulting formalism provides a unified and structurally consistent description of dissipative quantum--classical systems.

A distinctive feature of the present approach is its ability to accommodate both canonical and noncanonical variables. In particular, it allows for the treatment of systems such as the Bloch--Torrey model \cite{bloch1940nuclear,torre1956bloch}, where the dynamics involves both a quantum density matrix and a classical magnetization field. In this context, the hybrid entropy naturally decomposes into a classical contribution associated with the macroscopic degrees of freedom and a quantum contribution given by the von Neumann entropy, averaged over the classical distribution. This decomposition provides a clear physical interpretation of entropy production, which manifests as a relaxation at the quantum level and  diffusion at the classical level.

The formalism developed here provides a systematic route to derive physically relevant evolution equations. In particular, a thermodynamically consistent generalization of the  Lindblad master equation is produced within our hybrid metriplectic framework and, similarly,   a generalization the Bloch--Torrey equations are produced. Furthermore, the approach yields an effective Hamiltonian structure compatible with dissipative dynamics, thereby addressing the problem of reconciling energy conservation with dissipative  processes in hybrid systems.

The main result of this work is the construction of the  thermodynamically consistent hybrid metriplectic dynamics in which coupling arises solely through dissipative processes, and from which both the generalized Lindblad and Bloch–Torrey equations follow as special cases. Therefore, the paper is organized as follows. In Sec.~\ref{Metriplectic Formalism}, we review the metriplectic formalism and introduce the operator-based framework for quantum systems. In Sec.~\ref{Metriplectic formalism for quantum-classical hybrids system}, we construct the hybrid quantum--classical metriplectic structure and derive the corresponding evolution equations. In Sec.~\ref{Applications}, we present applications, including the derivation of the generalized  Lindblad  and generalized Bloch--Torrey systems. Finally, conclusions and perspectives are given in Sec.~\ref{conclusion}.

\section{Metriplectic Formalism}
\label{Metriplectic Formalism}

In this section, we briefly review the metriplectic formalism and introduce the operator-based framework required for the quantum and hybrid constructions that are developed later.

\subsection{Operator and functional calculus}

Let $\psi$ denote a dynamical variable which we assume lives in some function space $\mathcal{B}$.  A \emph{functional} refers to a mapping from  $\mathcal{B}$ into the field of real or complex numbers.  Here we will only consider those that map to real numbers.  Like ordinary functions, functionals possess notions of continuity and differentiability, as well as rules such as the chain rule and Leibniz rule. As in  classical dynamics, we assume that the system under consideration is contained in a domain $\Omega$ so $\psi\colon \Om\to \R$. The \emph{functional derivative} of $F\colon \mathcal{B}\to \R$, denoted by $\de F/\de \psi$,  is defined in terms of the directional derivative by
\bq
\delta F[\psi; \delta \psi] = \lim_{\varepsilon \rightarrow 0} \frac{F[\psi+\varepsilon\delta \psi] - F[\psi]}{\varepsilon}=\int_{\Om} \frac{\de F}{\de \psi} \de \psi\,,
\label{functdiff}
\eq
where for simplicity we drop the infinitesimal integration increment.  We shall assume that the limit exists, so the functional derivative,  $\de F/\de\psi$,  represents  an infinite-dimensional analog of the gradient. This fundamental notion is essential in Hamiltonian field theories as well as in the metriplectic formalism for continuous (infinite-dimensional) classical dynamics \cite{pjm82,pjm98,pjmC20,pjmZ25,pjmZB24,pjmBKM26}.

Since our goal is to develop a hybrid framework that combines classical and quantum mechanics, we will now introduce an analogous key tool for the space of operators.

Let $\mathbb{H}$ be a Hilbert space, denote by $\mathcal{L}(\mathbb{H})$ the set of linear operators defined on $\mathbb{H}$ and by $\mathcal{S}(\mathbb{H})$ the set of Hermitian operators. We define \emph{observables} in terms of   special kinds of functionals  that map $\mathcal{S}(\mathbb{H})$ to the real numbers, expectation values.  For example, for a linear operator $A\in \mathcal{S}(\mathbb{H})$ we denote the associated  observable by
\bq
\mathcal{A}[\rho] = \operatorname{Tr}(\rho A)\,. 
\eq
In finite dimensions $\rho$ and $H$ are matrices and $\operatorname{Tr}$ is the usual trace of a matrix product, while in infinite dimensions for mixed states the sums extend to infinity or they can be written in terms of integrals or the bra-ket notation. Thus, in all cases, observables are linear functionals of $\rho$.

Because observables are functionals, we can perform their functional differentiation just as in \eqref{functdiff}.
In this case the density \(\rho\) is varied slightly, namely
\[
\rho \to \rho + \varepsilon\, \eta,
\]
where \(\rho + \varepsilon\, \eta\) remains in the domain of definition. For observable $\mathcal{A}[\rho]$ we  have 
\bq
\mathcal{A}[\rho +\varepsilon \eta] = \mathcal{A}[\rho] + \varepsilon\operatorname{Tr}(\na_{\rho}\mathcal{A}\,\,\eta) + o(\varepsilon)\,,
\label{observable-derivative}
\eq
where we use the symbol $\na_{\rho}\mathcal{A}$ to denote the functional derivative of the observable $\cala$  in this quantum context. Here, we assumed that the limit $o(\varepsilon)/\varepsilon \to 0 \quad \text{as } \varepsilon \to 0 $. We will call  the quantity $\nabla_{\rho}\mathcal{A}$ the \emph{observable derivative} with respect the density $\rho$. Under this definition,  just as for \eqref{functdiff}, \(\nabla_{\rho}\mathcal{A}\) inherits the usual properties of derivatives: it is linear, satisfies the Leibniz rule for pointwise products of observables, and it obeys the chain rule for compositions with differentiable functions.

For clarity, we illustrate the preceding definitions with two standard examples from quantum mechanics. 
Let \(\rho\) denote the quantum density that describes  a mixed quantum state on  \(\mathbb{H}\), and let \(H\) be the Hamiltonian operator.
We consider two observables of fundamental physical importance: the expectation value of the Hamiltonian, which we interpret as the total energy, and the expectation value of \(-\ln \rho\), which corresponds to the total entropy, known as the von Neumann entropy.
These quantities are given by
\bq
\mathcal{H}(\rho) = \operatorname{Tr}(\rho H),
\qquad
\mathcal{S}(\rho) = -\operatorname{Tr}(\rho \ln \rho)\,.
\label{Tot-energy1}
\eq
Using the definition of the observable derivative, their derivatives with respect to \(\rho\) are readily obtained as
\bq
\nabla_{\rho}\mathcal{H} = H\,,
\qquad
\nabla_{\rho}\mathcal{S} = - \ln \rho - \mathbb{1}\,,
\eq
where \(\mathbb{1}\) denotes the identity operator on the Hilbert space \(\mathbb{H}\). Although, technically $\mathcal{S}(\rho)$ is not an observable, we use the $\nabla_{\rho}$ symbol for its functional derivative because it lies in the quantum sector.

\subsection{Metriplectic Framework}

\subsubsection{Poisson Brackets}

Nondissipative time evolution in Hamiltonian systems is generated via Poisson brackets acting on an appropriate Hamiltonian. The Poisson bracket is a bilinear operator, denoted 
$\{\,.\,,\,.\,\}$, characterized by antisymmetry, the Leibniz (product) rule, and the Jacobi identity. It acts on sufficiently regular real-valued functions, functionals, or observables defined on the space of independent variables describing a nonequilibrium system. These structural properties are well known to encode the essential features of mechanistically controlled nondissipative (ideal) dynamics of physical quantities. In particular, they ensure conservation of the energy and generate a consistent phase-space flow. For classical dynamics, a wide variety of Poisson brackets have been constructed for both finite- and infinite-dimensional systems and we refer the reader to the extensive literature for detailed examples and constructions \cite{pjmG80,pjm82,pjm98,pjmT00,hamdi,pjmDL16,pjmC20, pjmU24,pjmS24}.

In this section, we focus on the Poisson bracket that generates the Hamiltonian dynamics of a quantum system.
In contrast to classical systems, quantum mechanics naturally encodes Hamiltonian  dynamics through operator commutators $[\,.\,,\,.\,]$, which possess the required antisymmetry, bilinearity, and Jacobi identity.
Guided by these structural properties, we construct a Poisson bracket using the commutator and the observable derivative introduced above ~\eqref{observable-derivative}.

The appropriate Poisson bracket for a quantum system, defined on two observables \(\mathcal{A}\) and \(\mathcal{B}\), is given by
\bq
\{\mathcal{A}, \mathcal{B}\}
=
\frac{1}{i\hbar}
\operatorname{Tr}
\bigl(
\rho\,[\,\nabla_{\rho}\mathcal{A}\,,\nabla_{\rho}\mathcal{B}\,]
\bigr)\,,
\label{PB1}
\eq
where \(\hbar\) denotes the reduced Planck constant and \(i=\sqrt{-1}\) is the imaginary unit.

For an  arbitrary operator \(A\) on  \(\mathbb{H}\) with associated  observable (expectation value)
\bq
\mathcal{A}(\rho) = \operatorname{Tr}(\rho A).
\eq
{There are two possible ways to look at the evolution of a  quantum system:
\begin{itemize}
    \item The Schr\"odinger picture, where the density $\rho(t)$  changes in time.
    \item The Heisenberg picture, where the quantum density  is fixed and the operators $A:=A(t)$ change in time. 
\end{itemize}
In this study we adopt the Schr\"odinger picture.}
Thus, we restrict attention to the case where the operator \(A\) is time independent.

Upon substituting \(\mathcal{A}\) into the Poisson bracket \eqref{PB1}, its time evolution is governed by
\bq
\frac{\md\mathcal{A}}{\md t} = \{\mathcal{A}, \mathcal{H}\},
\label{Evolution1}
\eq
where \(\mathcal{H}\) denotes the total energy observable defined in \eqref{Tot-energy1}.
The left-hand side of \eqref{Evolution1} is given by
\bq
\frac{\md\mathcal{A}}{\md t}
=
\operatorname{Tr}\!\left(\frac{\md\rho}{\md t} A
\right),
\eq
while the right-hand side evaluates to
\[
\{\mathcal{A}, \mathcal{H}\}
= \frac{1}{i\hbar}\operatorname{Tr}(\rho\,[A,H])=
\operatorname{Tr}\!\left(
\frac{1}{i\hbar}[H,\rho]\,A
\right).
\]
Equating both expressions yields
\bq
\operatorname{Tr}\!\left(\frac{\md\rho}{\md t}\, A
\right)
=
\operatorname{Tr}\!\left(
\frac{1}{i\hbar}[H,\rho]\,A
\right).
\eq
Since the operator \(A\) is arbitrary in \(\mathcal{L}(\mathbb{H})\), we obtain the following evolution equation for a $\rho$ that describes a the mixed quantum state: 
\bq
\frac{\md\rho}{\md t}
=
\frac{1}{i\hbar}[H,\rho]\,.
\label{VNeq}
\eq
Equation \eqref{VNeq} is the von Neumann equation describing Hamiltonian  quantum dynamics. 

\emph{Casimir invariants} are  special observables \(\mathcal{C}\) that satisfy
\bq
\{\mathcal{C}, \mathcal{A}\} = 0,
\qquad
\forall\, \mathcal{A},
\eq
and are therefore constants of motion for any choice of Hamiltonian.
These  invariants reflect  degeneracies of the Poisson bracket and play a central role in enforcing constraints and thermodynamic consistency. For the quantum Poisson bracket defined in \eqref{PB1}, A Casimir invariant is any functional  \(\mathcal{S}\) whose observable derivative commutes with the density  \(\rho\),
\bq
\bigl[\nabla_{\rho}\mathcal{S},\, \rho\bigr] = 0\,.
\eq
Indeed, for an arbitrary observable \(\mathcal{A}\), we have
\bal
\{\mathcal{S}, \mathcal{A}\}&=
\frac{1}{i\hbar}
\operatorname{Tr}
\left(
\rho
\left[
\nabla_{\rho}\mathcal{S},
\nabla_{\rho}\mathcal{A}
\right]
\right)\nn\\
&=\frac{1}{i\hbar}
\operatorname{Tr}
\left(
\nabla_{\rho}\mathcal{A}
\left[
\rho,
\nabla_{\rho}\mathcal{S}
\right]
\right)=0\,,
\label{quanPB}
\eal
where the last equality follows from the assumed commutation of \(\nabla_{\rho}\mathcal{S}\) with \(\rho\) and the cyclicity of the trace. Classical examples of Casimir invariants in the quantum setting are some   commonly used entropy functionals: 
 the von Neumann entropy defined in \eqref{Tot-energy1}, the linear entropy,
\bq
\mathcal{S}_{l}(\rho) = 1 - \operatorname{Tr}(\rho^{2}),
\eq
and the so-called  Tsallis \(q\)-entropy,
\begin{equation}
    \mathcal{S}_{q}(\rho) = \frac{1}{q-1}\left(1 - \operatorname{Tr}(\rho^{q})\right),
\qquad q > 1,\label{eq:Tsalli}
\end{equation}
which has been extensively studied in quantum statistical mechanics \cite{tsallis2009nonadditive}. In the  limit $q \rightarrow 1$,  $\mathcal{S}_{q}$ becomes the von Neumann entropy \eqref{Tot-energy1}.

\subsubsection{Metriplectic 4-Bracket}
\label{sssec:M4B}

Metriplectic 4-brackets are employed to construct the dissipative contribution to the time evolution of functionals.
This framework, originally introduced in \cite{pjmU24} for classical dynamics and subsequently developed and extended in several works (e.g., \cite{pjmZB24,pjmZ25,pjmS25,pjmHB26}), provides a systematic and powerful tool for deriving thermodynamically consistent dynamical systems.
Within this approach, one can construct evolution equations that respect both the first law of thermodynamics, through energy conservation, and the second law, through  entropy production.
Moreover, these properties hold for arbitrary choices of Hamiltonian and entropy functionals, making the metriplectic 4-bracket framework particularly well suited for the formulation of general nonequilibrium dynamics. The metriplectic 4-bracket, which operates on four functionals,  denoted by \((\,\cdot\,,\,\cdot\,;\,\cdot\,,\,\cdot\,)\), is characterized by the following properties:

\smallskip
 
\noindent\emph{(i) Linearity in all arguments}: for any \(\lambda \in \mathbb{R}\),
\bq
(F + \lambda H, K; G, N)
=
(F, K; G, N)
+
\lambda (H, K; G, N).
\label{mullin}
\eq
\emph{(ii) Algebraic symmetries}:
\bal
(F, K; G, N) &= - (K, F; G, N)\,, \label{sym1} \\
(F, K; G, N) &= - (F, K; N, G)\,, \label{sym2} \\
(F, K; G, N) &= {(G, N; F, K)}\,. \label{sym3}
\eal
\emph{(iii) Derivation property in all arguments}: for example,
\bq
(FK, H; G, N)
=
F(H, K; G, N)
+
(F, K; G, N)H\,,
\label{derivp}
\eq
where $FK$ is pointwise multiplication. 

Both the multilinearity property of \eqref{mullin} and the derivation property  of \eqref{derivp} arise because the 4-bracket  is linearly dependent on the functional derivatives of each functional of  its arguments with respect to the dynamical variables (cf.\  the quantum Poisson bracket of \eqref{quanPB}). Instead of writing down a general 4-bracket expression that makes this manifest,  we refer the reader to \cite{pjmU24} for such an expression;  we will  show how it goes for particular examples in Sec.~\ref{Applications}.

One way to create a specific metriplectic 4-bracket  that has  the requisite symmetry  properties \eqref{sym1}--\eqref{sym3} is by using the Kulkarni-Nomizu (K-N) product \citep{Kulkarni,nomizu1971spaces}, denoted by $\KN$. (See also \cite{fiedler03} for relevant theorems.)  We define the 4-bracket in terms of the $\KN$ product of two symmetric bilinear operators, say $\Si$ and $M$, that
satisfy all of the axioms of an inner product space, except the non-degeneracy condition, as follows: 
\bal
(F, K ; G, N)&=(\Sigma \KN M)\left(F,K, G,N\right) 
\nn\\
&= \,\Sigma\left(F, G \right) M\left(K, N\right)
\nn\\
&\, -\Sigma\left(F , N \right) M\left(K , G \right) \nn\\
&\, +M\left(F , G \right) \Sigma\left(K , N \right)
\nn\\
&\,  -M\left(F , N \right) \Sigma\left(K , G \right)\,.
\eal

For a closed system, the total time evolution of any functional, say $F$,  can be written as the sum of its Hamiltonian  and dissipative contributions, generated respectively by the Poisson bracket and the metriplectic 4-bracket in the following manner:
\bq
\dot{F} = \{F,\mathcal{H}\} + (F,\mathcal{H};\mathcal{S},\mathcal{H})\,,
\eq
where $\mathcal{H}$ and $\mathcal{S}$ are the Hamiltonian and entropy, respectively, and  $\mathcal{S}$ is a chosen Casimir invariant. Thus,  automatically we have the  thermodynamic laws
\bal
\dot{\mathcal{H}} &= \{\mathcal{H},\mathcal{H}\} + (\mathcal{H},\mathcal{H};\mathcal{S},\mathcal{H}) = 0\,,
\\
\dot{\mathcal{S}} &= \{\mathcal{S},\mathcal{H}\} + (\mathcal{S},\mathcal{H};\mathcal{S},\mathcal{H}) = (\mathcal{S},\mathcal{H};\mathcal{S},\mathcal{H}) \geq 0\,.
\eal

However, many quantum systems of physical interest are inherently open, interacting with an external environment. In such cases, a purely quantum system cannot be described within the metriplectic framework. This limitation motivates the introduction of a quantum--classical hybrid  metriplectic formulation, in which the subsystem of interest is quantum while its surrounding environment is described classically. When the combined system is considered as a whole, its dynamics must remain thermodynamically consistent and can therefore be formulated within the metriplectic framework. The clarification of this approach and the corresponding dynamical picture is the subject of Sec.\ \ref{Metriplectic formalism for quantum-classical hybrids system}.

\section{Metriplectic formalism for quantum-classical hybrid system}
\label{Metriplectic formalism for quantum-classical hybrids system}

\subsection{Quantum-classical hybrid Poisson bracket }

We consider a hybrid setting in which a quantum subsystem interacts with a classical environment. 
Following earlier work \cite{elze2012linear,caro1999impediments,peres2001hybrid,buric2012statistical}, this problem may be treated within a linear framework by introducing canonical variables for each sector: {Hamiltonian phase-space variables \((x_k, p_k)\) for the classical system and $X_i$ and $P_i $ are the real  and  imaginary parts of the wave function in some chosen basis}. The phase space of the combined system is then taken as the Cartesian product state space that is 
canonically coordinated by $2(n + N)$-tuples \((x_k, p_k, X_i, P_i)\), which suggest a canonical Poisson bracket of the form
\bal
\{F,G\}_{\times}
&=
\{F,G\}_{\mathrm{CL}}
+
\{F,G\}_{\mathrm{QM}}\nn\\
&=
\sum_{k=1}^n
\left(
\frac{\partial F}{\partial x_k}
\frac{\partial G}{\partial p_k}
-
\frac{\partial F}{\partial p_k}
\frac{\partial G}{\partial x_k}
\right)
\nn\\
&\hspace{1.0cm}+\sum_{i=1}^N
\left(
\frac{\partial F}{\partial X_i}
\frac{\partial G}{\partial P_i}
-
\frac{\partial F}{\partial P_i}
\frac{\partial G}{\partial X_i}
\right).
\eal

Within this formulation,  in the absence of interactions between the classical and quantum sectors, the dynamics is generated by a separable total Hamiltonian of the following form: 
\bq
H_{\Si}(x_k,p_k; X_i,P_i)
:=
H_{\mathrm{CL}}(x_k,p_k)
+
H_{\mathrm{QM}}(X_i,P_i)\,,
\eq
where \(H_{\mathrm{CL}}\) denotes the Hamiltonian of the classical system and \(H_{\mathrm{QM}}\) the Hamiltonian associated with the quantum subsystem.
This additive structure reflects the dynamical separability of the two sectors in the absence of coupling.

In the present work, however, the variables employed in our formalism are not canonical and, consequently, the associated Poisson bracket does not have  canonical form. Nevertheless, we retain a linear structure and construct the hybrid Poisson bracket as a direct sum of the classical and quantum contributions.

Furthermore, we restrict attention to the case in which nondissipative  interactions between the classical and quantum sectors are neglected. Couplings between the two subsystems are instead introduced exclusively through the dissipative part of the dynamics, whose structure and physical interpretation will be discussed in detail in Sec. \ref{Quantum-classical hybrids Metriplectic 4-bracket}. 

We  assume that any functional \(\mathcal{F}\) of the hybrid system admits a decomposition into a classical and a quantum contribution, i.e., 
\bq
F := F^{q}+ F^{c}\,.
\label{decomp}
\eq
where typically $F^c$ will be a quantum observable.
This means that the space of functionals \(\mathfrak{F}\) is the direct sum of the subspace of classical observables \(\mathfrak{C}\) and the subspace of quantum observables \(\mathfrak{Q}\), namely
\bq
\mathfrak{F} = \mathfrak{C} \oplus \mathfrak{Q}\,.
\eq

Also within our framework, the Poisson bracket of two arbitrary functionals \(F\) and \(G\) is defined as the direct sum of their respective classical and quantum brackets,
\bq
\{F,G\}_{h}
=
\{F^{c}, G^{c}\}_{c}
+
\{F^{q}, G^{q}\}_{q}\,,
\eq
where $\{F,G\}_{h}$ generates the nondissipative dynamics, with   \(\{\,\cdot\,,\,\cdot\,\}_{c}\) generating  the Hamiltonian  classical dynamics and \(\{\,\cdot\,,\,\cdot\,\}_{q}\) generating  the Hamiltonian quantum dynamics. The hybrid bracket $\{\cdot,\cdot\}_h$ inherits a bilinear structure, and satisfies the following properties:
\bal
\{F^c + F^q , G\}_h &= \{F^c,G\}_h + \{F^q,G\}_h\,,
\\
\{F^c , G^c + G^q\}_h &=    \{F^c,G^c\}_c + \{0,G^q\}_q 
\nn\\
&= \{F^c,G^c\}_c\,,
\\
\{F^q , G^c + G^q\}_h &= \{0,G^c\}_c + \{F^q,G^q\}_q
\nn\\
&= \{F^q,G^q\}_q\,.
\eal

Consistently, the hybrid Hamiltonian of the system is taken to be a sum of the classical Hamiltonian functional \(\mathcal{H}^{c}\) and the quantum Hamiltonian \(\mathcal{H}^{q}\),
\bq
\mathcal{H} :=\mathcal{H}^{c}+ \mathcal{H}^{q},
\eq
with an analogous construction assumed for the hybrid entropy.
As a consequence, the nondissipative time evolution of an functional $F$ is governed by
\bq
\dot{F}
=
\{F, \mathcal{H}\}_{h}
=
\{F^{c}, \mathcal{H}^{c}\}_{c}
+
\{F^{q}, \mathcal{H}^{q}\}_{q},
\eq
which explicitly reflects the separability of the classical and quantum sectors in the absence of Hamiltonian  coupling.

\subsection{Quantum-classical hybrid Metriplectic 4-bracket}
\label{Quantum-classical hybrids Metriplectic 4-bracket}

When no interactions are present between a quantum subsystem and its classical environment,  the dynamics of each subsystem may be formulated independently.
In this decoupled setting, the time evolution of classical and quantum functionals would  be  given by something like 
\bal
\dot{F}^{c}
&=
\{F^{c}, \mathcal{H}^{c}\}_{c}
+
(F^c,\mathcal{H};\mathcal{S},\mathcal{H})_c\,,
\label{no-interaction-quantum}
\\
\dot{F}^{q}
&=
\{F^{q}, \mathcal{H}^{q}\}_{q}
+
(F^q,\mathcal{H};\mathcal{S},\mathcal{H})_q\,,
\label{no-interaction-classic}
\eal
where recall  \(\{\,\cdot\,,\,\cdot\,\}_{c}\) and \(\{\,\cdot\,,\,\cdot\,\}_{q}\) denote the classical and quantum Poisson brackets, respectively, and it remains to define what is meant by $(F^c,\mathcal{H};\mathcal{S},\mathcal{H})_c$ and  $(F^q,\mathcal{H};\mathcal{S},\mathcal{H})_q$,  which describe  the dissipative dynamics of each subsystem.

Thermodynamic consistency could be  ensured separately for each subsystem. However, when interactions are present—so that the quantum subsystem is open to the classical environment and vice versa—this decoupled structure no longer holds.
In particular, one cannot generally expect the conservation of energy of each subsystem, which would require
\bq
(\mathcal{H}^c,\mathcal{H};\mathcal{S},\mathcal{H})_c = 0
\quad\mathrm{and}\quad
(\mathcal{H}^q,\mathcal{H};\mathcal{S},\mathcal{H})_q = 0\,.
\eq
Moreover, unlike the Poisson brackets, the dissipative  contributions cannot simply be added without violating fundamental thermodynamic constraints.
Indeed, in general,
\bal
\dot{\mathcal{H}} &=\dot{\mathcal{H}^c}+ \dot{\mathcal{H}^q}
\nn\\
&=(\mathcal{H}^c,\mathcal{H};\mathcal{S},\mathcal{H})_c
+
(\mathcal{H}^q,\mathcal{H};\mathcal{S},\mathcal{H})_q
\neq 0\,,
\eal
so that a naive summation of these two  quantities would fail to guarantee conservation of the total energy.

These considerations motivate the introduction of a hybrid metriplectic 4-bracket, which in fact is {\bf not} simply the sum of classical and quantum components.
Using the two symmetric operators \(M\) and \(\Sigma\), akin to those of Sec.~\ref{sssec:M4B}, together with their Kulkarni--Nomizu product, we construct a hybrid 4-bracket that generates the dissipative dynamics of the coupled quantum--classical system.
It is defined as follows:
\begin{align}
(F,K;G,N)_{h}
=&
\quad\Sigma(F^{q}, G^{q})\, M(K^{c}, N^{c})
\nn\\
&-
\Sigma(F^{q}, N^{q})\, M(K^{c}, G^{c})
\nonumber\\
&
+
M(F^{c}, G^{c})\, \Sigma(K^{q}, N^{q})
\nn\\
&-
M(F^{c}, N^{c})\, \Sigma(K^{q}, G^{q})\,.
\label{h4bkt}
\end{align}
This 4-bracket does indeed  satisfy the defining properties of metriplectic 4-brackets \eqref{sym1}--\eqref{sym3}. {Furthermore, the hybrid bracket enjoys the following properties:
\bal
(F^c + F^q , K;G,N)_h &= (F^c,K;G,N)_h
\label{prop1}\\
&\hspace{.75cm}  + (F^q,K;G,N)_h \,,
\nn\\
(F^c,K;G,N)_{h}&= M(F^{c}, G^{c})\, \Sigma(K^{q}, N^{q})
\label{prop3}\\
&\hspace{.75cm} -
M(F^{c}, N^{c})\, \Sigma(K^{q}, G^{q})\,,
\nn\\
(F^q,K;G,N)_{h}
&= \Sigma(F^{q}, G^{q})\, M(K^{c}, N^{c})
\label{prop2}\\
&\hspace{.75cm} -
\Sigma(F^{q}, N^{q})\, M(K^{c}, G^{c})
\,.
\nn
\eal

With the hybrid Poisson bracket governing the Hamiltonian  dynamics and the hybrid metriplectic 4-bracket of \eqref{h4bkt} generating dissipative coupling, the time evolution of any functional  \(F\) of the combined classical--quantum system takes the form
\begin{equation}
\dot{F}
=
\{ F, \mathcal{H} \}_h
+
(F, \mathcal{H}; \mathcal{S}, \mathcal{H})_h\,,
\end{equation}
where \(\mathcal{H}\) denotes the total Hamiltonian of the hybrid system,
\begin{equation}
\mathcal{H}
:=\mathcal{H}^{c}+ \mathcal{H}^{q},
\end{equation}
and \(\mathcal{S}\) is the total entropy,
\begin{equation}
\mathcal{S}
:=\mathcal{S}^{c}+ \mathcal{S}^{q}.
\end{equation}
Given the properties \eqref{prop1}--\eqref{prop3} we can obtain the dynamics of any functional of the form of \eqref{decomp}:
\bal 
\dot{F^c}&=\{F^c,\calh^c\}_c + (F^c,\calh;\cals,\calh)_{h}
\nn\\
&= \{F^c,\calh^c\}_c  +M(F^{c}, \cals^{c})\, \Sigma(\calh^{q}, \calh^{q})
\nn\\
&\hspace{3cm}
- M(F^{c}, \calh^{c})\, \Sigma(\calh^{q}, \cals^{q})
\label{dotFc}\\
\dot{F}^q&=\{F^q,\calh^q\}_q + (F^q,\calh;\cals,\calh)_{h}
\nn\\
&= \{F^q,\calh^q\}_q  + \Sigma(F^{q}, \cals^{q}) M(\calh^{q}, \calh^{q})\, 
\nn\\
&\hspace{3cm} 
- \Sigma(F^{q}, \calh^{q}) M(\calh^{c}, \cals^{c})\,.
\label{dotFq}
\eal

By construction, the  Poisson bracket dynamics conserves the total Hamiltonian, while the hybrid metriplectic 4-bracket is degenerate with respect to \(\mathcal{H}\) and produces a non-negative entropy production.
Consequently, the full hybrid system satisfies both the first law of thermodynamics, through conservation of energy, and the second law, through the monotonic increase of the total entropy.
The coupled quantum--classical hybrid dynamics is therefore thermodynamically consistent:
\bq
\dot{\mathcal H}
=
\{\mathcal H,\mathcal H\}_h
+
(\mathcal H,\mathcal H;\mathcal S,\mathcal H)_h = 0,
\eq
\bq
\dot{\mathcal S}
=(\mathcal S,\mathcal H;\mathcal S,\mathcal H)_h \geq 0.
\eq

The  metriplectic framework relies crucially on degeneracy conditions that guarantee thermodynamic consistency. In the absence of interactions between subsystems, a strong form of degeneracy is typically imposed, namely that the dissipative  operators annihilate the Hamiltonian. Concretely, for the classical and quantum subsystems this would  require  
\bq
M(F^c, \calh^c)= 0
\quad\mathrm{and}\quad 
\Si(F^q,\calh^q)= 0\,,
\eq
for any purely classical functional  $F^{c}$ and any quantum observable $F^{q}$. These conditions ensure that dissipation does not affect the  energy conserving Hamiltonian dynamics of either subsystem (cf.\ \eqref{dotFc} and \eqref{dotFq}).

In the interacting setting considered here, however, the decoupled evolution equations of the form of \eqref{no-interaction-quantum}  and \eqref{no-interaction-classic} no longer describe the full dynamics. In particular, there exist hybrid functionals  \(F =F^{q}+F^{c}\) for which
\bq
M(F^{c}, \mathcal{H}^{c}) \neq 0
\qquad\mathrm{and\qquad}
\Si(F^{q}, \mathcal{H}^{q}) \neq 0\,.
\eq
We assume that the classical environment admits a fixed Hamiltonian \(\mathcal{H}^{c}\) such that $M(\mathcal{H}^{c}, \mathcal{H}^{c}) \neq 0$. Without loss of generality, we impose the normalization
\bq
M(\mathcal{H}^{c}, \mathcal{H}^{c}) \equiv 1\,.
\label{normM}
\eq
This assumption allows us to precisely characterize when the quantum subsystem remains unaffected by dissipative  classical influences.   

\begin{definition}\label{definition1}
The quantum subsystem is said to be \emph{unaffected by the classical environment} if no entropy production or dissipative contribution originating from the classical sector influences the quantum dynamics. Equivalently, for any quantum observable \(F^{q}\), the dissipative  contribution satisfies
\bq
(F^{q}, \mathcal{H}; \mathcal{S}, \mathcal{H})_h
=
\Si(F^{q}, \mathcal{S}^{q}),
\eq
so that only the intrinsic quantum dissipation acts on the quantum subsystem.
\end{definition}

\begin{lemma}\label{lemma1}
The condition
\[
(F^{q}, \mathcal{H}; \mathcal{S}, \mathcal{H})_h
=
\Sigma(F^{q}, \mathcal{S}^{q})
\quad \text{for all } F^{q} 
\]
holds if and only if
\[
(F^{q}, \mathcal{H}; \mathcal{S}^{c}, \mathcal{H})_h = 0
\quad \text{for all } F^{q},
\]
which is in turn equivalent to the degeneracy condition
\[
M(\mathcal{S}^{c}, \mathcal{H}^{c}) = 0\,.
\]
\end{lemma}

In this setting, we now rewrite the equations governing the time evolution of observables within the hybrid metriplectic framework. We begin by considering a purely quantum observable 
\(F^{q}\), meaning that its classical component vanishes identically, i.e., $F^c\equiv0$.  Then, \eqref{dotFq} for the evolution equation of an arbitrary quantum observable simplifies to
\begin{equation}
\dot{F^{q}}
=
\{F^{q}, \mathcal{H}^{q}\}_{q}
+
\Sigma(F^{q}, \mathcal{S}^{q})\,,
\label{general-evolution-quatum2}
\end{equation}
where we have invoked \eqref{normM}.
This expression describes a quantum system undergoing  Hamiltonian evolution generated by \(\mathcal{H}^{q}\), together with dissipative dynamics driven by the quantum entropy \(\mathcal{S}^{q}\).

A completely analogous calculation can be carried out for a purely classical functional $F^{c}$, for which $F^{q} \equiv 0\).
Its time evolution according to \eqref{dotFc} is given by
\bal
\dot{F^{c}}
&=
\{F^{c}, \mathcal{H}^{c}\}_{c}
\label{dotFC}\\
&\quad + M(F^{c}, \mathcal{S}^{c})\, \Sigma(\mathcal{H}^{q}, \mathcal{H}^{q})
-
M(F^{c}, \mathcal{H}^{c})\, \Sigma(\mathcal{H}^{q}, \mathcal{S}^{q})\,.
\nn
\eal
 Equation \eqref{dotFC}  makes  explicit how the dissipative  quantum sector contributes to the effective dissipative dynamics of the classical environment.

In the next section \ref{Applications}, this formalism will be employed to derive concrete dynamical equations of physical relevance. In particular, we show how the present hybrid metriplectic approach leads naturally to the Lindblad master equation that describe the open quantum systems in Sec. \ref{Lindblad master equation}, as well as to the Bloch--Torrey equation as a paradigmatic example of quantum–classical hybrid dynamics in Sec. \ref{Bloch-Torrey equation}.

\section{Applications}
\label{Applications}

{The purpose of this section is to illustrate how the general metriplectic framework developed in the previous sections can be applied to concrete physical models. 
We begin with purely quantum systems, where the formalism naturally recovers well-known dissipative dynamics, and then progressively extend the analysis to more structured settings, including energy-conserving systems and hybrid quantum--classical models.

We start by considering open quantum systems, for which dissipation and irreversibility play a fundamental role. In this context, the metriplectic structure provides a natural framework to recover the Lindblad master equation.}

\subsection{Lindblad Master equation}
\label{Lindblad master equation}

A central challenge in the theoretical description of open quantum systems is the incorporation of environmental effects into the quantum dynamics in a manner that preserves physical consistency. A particularly effective approach is based on the quantum density formalism and its associated master equations. While formally equivalent to the Schr\"odinger and Heisenberg pictures for closed systems, the um quantdensity  framework naturally accommodates incoherent processes arising from system–environment interactions. In particular, the density  allows for the representation of statistical mixtures of quantum states, a feature that is essential for modeling irreversible phenomena such as decoherence and dissipation. For these reasons, master equation approaches provide a natural bridge between microscopic quantum dynamics and nonequilibrium thermodynamic descriptions.

The most general form of a Markovian, completely positive, trace-preserving evolution for the density matrix was independently derived in the seminal works of Lindblad \cite{lindblad1976generators} and Gorini, Kossakowski, and Sudarshan \cite{gorini1976completely} (GKSL). The resulting evolution equation, commonly referred to as the GKSL  master equation, governs the time evolution of the system quantum density  \(\rho(t)\) and can be written, in diagonal form, as
\begin{equation}
\frac{\md \rho}{\md t}
=
\frac{1}{i\hbar}[H,\rho]
+
\sum_{k} \gamma_k
\left(
L_k \rho L_k^\dagger
-
\frac{1}{2}[L_k^\dagger L_k,\rho]_s
\right),
\end{equation}
where $[\, , \, ]$ is the operator commutator,  \(H = H^\dagger\) is the system Hamiltonian and \(\{L_k\}\) are the Lindblad, or jump, operators encoding the dissipative influence of the environment. The coefficients \(\gamma_k \geq 0\) are damping rates, and the anticommutator is defined by \([A,B]_s = AB + BA\).

In the present work, we restrict attention to the case of Hermitian Lindblad operators, \(L_k^\dagger = L_k\). Under this assumption, the dissipative contribution simplifies and the master equation reduces to 
\begin{equation}
\frac{\md \rho}{\md t}
=
\frac{1}{{i}\hbar}[H,\rho]
-
\sum_{k} \frac{\gamma_k}{2}
\,\big[L_k,[L_k,\rho]\big]\,.
\end{equation}
Note, with slight modification our results can be adapted for anti-Hermitian operators.

In order to connect the general formalism with the standard description of open quantum systems, we now specialize the construction of the symmetric operator $\Si$ and show how the Lindblad dynamics emerges.

\subsubsection{Open quantum systems}

To derive the Lindblad master equation within the metriplectic formulation, which recall describes the dynamics of an open quantum subsystem embedded in a classical environment, we specify the structure of the symmetric operator \(\Sigma\) (see Sec.~\ref{Quantum-classical hybrids Metriplectic 4-bracket}). This structure is based on the Bogoliubov inner product, which appears  in the context of quantum statistical mechanics (see, e.g., \cite{petz1993bogoliubov}).

Let \(R\) be a self-adjoint operator on the Hilbert space 
\(\mathbb{H}\). The Bogoliubov inner product (also referred to as the Duhamel two-point function, the Kubo--Mori--Bogoliubov inner product, or the Bogoliubov scalar product) between two operators \(X\) and \(Y\) is defined by
\begin{equation}
\langle X , Y \rangle_{R}
=
\int_{0}^{1}
\operatorname{Tr}
\!\left(
\mathrm{e}^{\lambda R}
X^{\dagger}
\mathrm{e}^{(1-\lambda)R}
Y
\right)
\,\md \lambda \,.
\end{equation}

In the present setting, we take \(R = \ln \rho\), which gives
\begin{equation}
\langle X , Y \rangle_{\ln \rho}
=
\int_{0}^{1}
\operatorname{Tr}
\!\left(
\rho^{\lambda}
X^{\dagger}
\rho^{1-\lambda}
Y
\right)
\,\md \lambda \,.
\label{BIP}
\end{equation}

Using the Bogoliubov inner product, we define the symmetric bilinear operator \(\Sigma\) acting on quantum observables \(F^{q}\) and \(G^{q}\) as
\begin{equation}
\Sigma(F^{q}, G^{q})
=
\sum_{k}
\frac{\gamma_k}{2}
\,
\left\langle
\big[ L_k , \nabla_{\rho} F^{q} \big],
\big[ L_k , \nabla_{\rho} G^{q} \big]
\right\rangle_{\ln \rho}\,,
\label{SLIP}
\end{equation}
where \(\{L_k\}\) are Hermitian operators and \(\gamma_k \geq 0\) are real coefficients. By construction, \(\Sigma\) is symmetric and positive semidefinite.

According to Definition~\ref{definition1} and the result of Lemma~\ref{lemma1}, we assume that the quantum subsystem is unaffected by the classical environment. Consequently, one may choose a symmetric positive semidefinite operator \(M\) such that
\bq
M(\mathcal{S}^{c}, \mathcal{H}^{c}) = 0\,,
\qquad
M(\mathcal{H}^{c}, \mathcal{H}^{c}) = 1 \,.
\eq
Under this assumption, the general evolution equation for a quantum observable \(F^{q}\),
\begin{equation}
\dot{F^{q}}
=
\{F^{q}, \mathcal{H}\}_{h}
+
(F^{q}, \mathcal{H}; \mathcal{S}, \mathcal{H} )_{h}\, ,
\label{general-evolution-quatum1}
\end{equation}
reduces to Eq.~\eqref{general-evolution-quatum2}.

We now consider the Poisson bracket \eqref{PB1} and insert the quantum Hamiltonian
\bq
\mathcal{H}^{q} = \operatorname{Tr}(\rho H),
\eq
together with the von Neumann entropy
\bq
\mathcal{S}^{q} = -\operatorname{Tr}(\rho \ln \rho),
\eq
as defined in \eqref{Tot-energy1}. For an observable of the form \(\mathcal{A} = \operatorname{Tr}(\rho A)\), with \(A \in \mathcal{L}(\mathbb{H})\), the evolution equation becomes
\begin{equation}
 \dot{\mathcal{A}}
=
\operatorname{Tr}\!\left(\frac{1}{{i}\hbar}[H,\rho]\,A\right)
+
\Sigma(\mathcal{A}, \mathcal{S}^{q}).
\end{equation}

According to the calculations presented in Appendix \ref{Appendix A}, the second term on the right-hand side can be evaluated explicitly. Consequently, the equation reduces to
\bal
\operatorname{Tr}\! \left(\frac{\md \rho}{\md t}\, A\right)
&=
\operatorname{Tr}\!\left(\frac{1}{{i}\hbar}[H,\rho]\,A\right)
\nn\\
&\qquad  -
\sum_{k} \frac{\gamma_k}{2}
\operatorname{Tr}
\!\left(\big[L_k,[L_k,\rho]\big]\, A
\right),
\eal
for arbitrary \(A \in \mathcal{L}(\mathbb{H})\). Since this equality holds for all such operators \(\cala\), it follows that the density operator \(\rho\) satisfies the Lindblad master equation
\bal
\frac{\md \rho}{\md t}
=
\frac{1}{{i}\hbar}[H,\rho]
+
\sum_{k}
\gamma_k
\left(
L_k \rho L_k
-
\frac{1}{2}
[L_k L_k,\rho]_s
\right).
\eal

{While the above discussion focused on general dissipative quantum dynamics, it is also of interest to identify situations in which the quantum subsystem preserves its energy. In Sec.~\ref{LBenergy} we investigate how this additional constraint restricts the structure of the Lindblad operators.}

\subsubsection{Energy-Conserving Quantum Systems }
\label{LBenergy}

{To assure energy conservation,  we follow the approach developed in~\cite{2017lindbladian}. 
In each case under consideration, we assume that the Hamiltonian operator depends explicitly on time. 
Our goal is to determine the resulting form of the Lindblad master equation under a set of hypotheses.

Let $\mathfrak{g}$ denote the smallest Lie algebra such that $H(t)\in\mathfrak{g}$ for all times $t$. 
In practice, the choice of $\mathfrak{g}$ is dictated by convenience and by the symmetries of the problem at hand.
We impose the following assumptions:
\begin{itemize}
    \item \textbf{$\bfH_1$:} The Lindblad operators $L_k$ are Hermitian elements of the Lie algebra $\mathfrak{g}$, i.e.\ $L_k^\dagger = L_k$, and are specified by the physical setting under study.
    \item \textbf{$\bfH_2$:} The quantum subsystem satisfies energy conservation,
    \begin{equation}
        \frac{\mathrm d \mathcal{H}}{\mathrm d t}
        = \frac{\mathrm d}{\mathrm d t} \operatorname{Tr}( \rho H)
        \equiv 0 .
        \label{dH^q/dtequiv0}
    \end{equation}
\end{itemize}
Under these restrictions, we show that the Lindblad master equation acquires an explicit and constrained structure. 

Thus, within the metriplectic framework $\calh$ satisfies
\begin{align}
\frac{\mathrm d \mathcal{H}}{\mathrm d t}
&=
\operatorname{Tr}\!\left(\rho\,\frac{\mathrm d H}{\mathrm d t}\right)
+
\operatorname{Tr}\!\left(\frac{\mathrm d \rho}{\mathrm d t}\, H\right)
\nonumber\\
&=
\operatorname{Tr}\!\left(\rho\, \frac{\mathrm d H}{\mathrm d t}\right)
+
\{\mathcal{H},\mathcal{H}\}
+
\Sigma(\mathcal{H},\mathcal{S})\,,
\end{align}
where $\cals$ is a quantum entropy functional.
Since the Poisson bracket vanishes identically, energy conservation is ensured if
\[
\operatorname{Tr}\!\left(\rho\, \frac{\mathrm d H}{\mathrm d t}\right)
+
\Sigma(\mathcal{H},\mathcal{S})
=0.
\]
Using the explicit form of the symmetric operator $\Sigma$, this condition becomes
\[
\operatorname{Tr}\!\left(\rho\frac{\mathrm d H}{\mathrm d t}\right)
-
\sum_k \gamma_k
\operatorname{Tr}\!\left(\rho\,\big[L_k,[L_k,H]\big]\right)
=0\,.
\]
A sufficient condition for~\eqref{dH^q/dtequiv0} to hold for arbitrary $\rho$ is therefore
\bq
\frac{\mathrm d H}{\mathrm d t}
-
\sum_k \gamma_k \big[L_k,[L_k,H]\big]
=0\,.
\eq

}

{
In all the following examples, we restrict our attention to the case $\gamma_1 = 1$ and $L_1 = L$, while all other coefficients vanish, i.e., $\gamma_k = 0$ for $k \neq 1$. Under this assumption, equation \eqref{dH^q/dtequiv0} reduces to
 \begin{equation}\label{eq:coefficents}
     \Dt{H} = \big[L, [L, H]\big]\,.
 \end{equation}
Let the Lie algebra  $\mathfrak{g}$ be generated by some basis of operators  $A_i$, with the associated structure constants $c_{ij}^k$, i.e., 
\begin{equation}
    [A_i, A_j] = c_{ij}^k\, A_k\,,
\end{equation}
were  repeated index summation convention is used.  Recall 
 structure constants $c_{ij}^k$  have the  antisymmetry conditions 
\bq
c_{ij}^k = - c_{ji}^k\,,
\eq 
and the Jacobi identity 
\bq
c_{im}^n c_{jk}^m + c_{jm}^nc_{ki}^m + c_{km}^nc_{ij}^m = 0\,.
\eq
In this basis, there are coefficients $\alpha_i(t) $ such that 
\bq
H(t) = \alpha^i(t) A_i \,.
\label{alpha}
\eq

In order to construct the master equation for  $\rho$ we need to  determine the coefficients $\beta^i$ such that 
\bq
L = \beta^i A_i\,.
\label{beta}
\eq
Using \eqref{alpha} and \eqref{beta} in  \eqref{eq:coefficents} we obtain  
\bq
\dot\alpha^m A_m = \big[\beta^iA_i, [\beta^j A_j, \alpha^k A_k] \big] 
\eq
which implies
\begin{equation}\label{eq:to_solve}
     \dot\alpha^m = \alpha^k\beta^i\beta^j c_{kj}^nc_{in}^m.
\end{equation}

}

\subsection{Examples}
{To illustrate the abstract results obtained in Sec.~\ref{Lindblad master equation}, we now consider several explicit examples. These examples demonstrate how the imposed structural and physical constraints translate into concrete forms of the Lindblad operators and the associated dynamics.}

\subsubsection{ Quantum harmonic oscillator}

{The quantum harmonic oscillator constitutes a fundamental model in both classical and quantum physics, describing systems subject to a restoring force proportional to the displacement. In the present setting, we consider a time-dependent harmonic oscillator, where the evolution is governed by a Hamiltonian whose stiffness coefficient varies in time. Thus the Hamiltonian is given by 
\begin{equation}
    H(t) = \frac{p^2 }{2} + \frac{k(t)}{2} x^2 \,,
\end{equation}
where $x$ and $p$ denote, respectively, the position and momentum operators, which are  linked by the Heisenberg identity 
\begin{equation}
    [x, p ] = i\hbar \,.
\end{equation}

Upon defining the operators 
\bq
    K_1 = \frac{p^2}{2}, \quad K_2 = \frac{x^2}{2}, \quad K_3 = \frac{xp + px}{2}\,,
    \label{k1_k2_k3}
\eq
 the commutation relations become 
\bal
    [K_1, K_2] &= -i\hbar K_3\,,\quad [K_2, K_3] = 2i\hbar K_2\,,
     \nn\\
     [K_3, K_1] &= 2i\hbar K_1\,.
\eal
This Lie algebra $\mathfrak{g}$, which is  generated by $(K_1, K_2, K_3)$, is  $\mathfrak{sl}(2,\C)$.  

The Hamiltonian in this basis is 
\bq
H(t) = K_1 + k(t) K_2,
\eq
so that $H(t) \in \mathfrak{g}$. Moreover we will assume that all the Lindblad operators are elements of $ \mathfrak{g}$, i.e., 
\bq
L = \beta_1 K_1 + \beta_2K_2 + \beta_3 K_3
,.
\eq
Thus, according to equation \eqref{eq:to_solve}, we obtain the following system:
\bq
\begin{cases}
    k(\beta_1)^2 - \beta_1\beta_2 +2 (\beta_3)^2 = 0 \\
    k \beta_1\beta_2 -(\beta_2)^2 - 2 k (\beta_3)^2 = \dot k/({2\hbar}) \\
    (k\beta_1 + \beta_2) \beta_3 = 0\,.
\end{cases}\label{k=cste}
\eq

From  the  equations  \eqref{k=cste}, we see the following possibilities:
\begin{enumerate}[(i)]
    \item  $\beta_1 = \beta_3 = 0$ and $(\beta_2)^2 = -\dot{k}(t)/2\hbar$.  Upon assuming  $\dot k(t) \leq 0 $, the Lindblad operator reads,
\bq
L = \sqrt{\frac{-\dot{k}(t)}{2\hbar}}\, K_2\,,
\eq
and the  master equation becomes
\begin{equation}
    i\hbar \Dt{\rho} = [H(t), \rho] 
    + i \frac{\dot k(t)}{8\hbar } \big[x^2 , [x^2, \rho]\big]\,.
    \label{case1} 
\end{equation}
\item  $\beta_3 = 0$ and  $\beta_2 = k \beta_1$. This case requires  $\dot k = 0 $.  With $k$ being  a constant  $L $ is merely  a multiple of the Hamiltonian,  i.e., 
\bq
L = \beta_1(t) H 
\label{case2L}
\eq
and  the master equation becomes 
\begin{equation}
    i\hbar \Dt{\rho} = [H(t), \rho] 
    -  i \beta_1(t) \big[H , [H, \rho]\big]\,.
    \label{case2} 
\end{equation}
\item $k\beta_1 + \beta_2 = 0$, leads to $k (\beta_1)^2 + (\beta_3)^2 = 0$ and $\dot k = 0$.
Hence, non-trivial solutions exist only if $k\leq 0$, in which case
\[
\beta_3 = \pm \sqrt{-k}\, \beta_1 .
\]
and the Lindblad operator has two possibilities, which we write as 
\bq
L_\pm = \beta_1(K_1 - k K_2 \pm   \sqrt{-k}\,  K_3)\,.
\label{LO3}
\eq
Under these assumptions, we obtain the following master equation:
\bq
    i\hbar \Dt{\rho} = [H(t), \rho]  
    - i \beta_1(t)  \big[K_\pm , [K_\pm, \rho]\big]\,, 
    \label{case3} 
\eq
where 
\[
K_\pm\coloneqq K_1 - k K_2 \pm \sqrt{-k} \,K_3\,.
\label{Kpm}
\]
Upon substituting the expressions for  $K_1$, $K_2$, and $K_3$ given in \eqref{k1_k2_k3}, this becomes
\bq
K_\pm= p^2 - k x^2 \pm \sqrt{-k}\,(xp+px)\,.
\eq
\end{enumerate}
}

In summary, the metriplectic formalism has produced  the master equations of   \eqref{case1}, \eqref{case2}, and \eqref{case3}.  Equation \eqref{case1} appeared in \cite{2017lindbladian}, while the other two appear to be new.  By construction all these equations are thermodynamically consistent,  preserving the energy and producing entropy, but the manner in which energy is redistributed between the quantum sector and the environment needs further elucidation.  

In  \cite{2017lindbladian} it was noted that  the master equation  \eqref{case1} of case(i) with   $\dot{k}\leq 0$   is consistent with $\dot\cals\geq 0$, and this gives rise to  flattening of the harmonic potential.  A flatter  potential  has a lower energy spectrum and energy conservation is maintained  because  the  oscillator absorbs  energy from the environment. 

For the master equation  \eqref{case2} of case (ii) we have the  proportionality  of  \eqref{case2L}. Consequently, 
the environment via $L$ continuously tracks the energy of the oscillator; however,   populations of the energy eigenbasis are unchanged so  no energy is exchanged with the environment.  The dissipative term  has the form of an energy-dephasing channel that  suppresses coherences in the energy eigenbasis.  Even though  no energy is exchanged with the environment,  entropy increases through the loss of quantum coherence.

The master equation  \eqref{case3} of case (iii)  is more interesting.
Since $H=K_1+ k K_2$,  the Lindblad operator of \eqref{LO3} is not proportional to the Hamiltonian.
Instead,  it is a particular linear combination of kinetic energy, potential energy and the dilation (squeezing) generator
$K_3= (xp +px)/2$.  
Notice that $K_3$ generates scale transformations (squeezing) in phase space: $x\mapsto e^{-\la} x, p\mapsto e^{\la} p$. 
Therefore, it  naturally entails   squeezing-like degrees of freedom, rather than just energy.  Thus, the environment does not monitor the energy,  but a specific observable belonging to the $\mathfrak{sl}(2)$  algebra.   The corresponding dissipative dynamics therefore acts on the phase-space geometry of the oscillator, damping particular squeezed combinations of position and momentum rather than the energy itself.  The full ramifications of this will be considered elsewhere.

\subsubsection{ Charged particle in Magnetic Field }

A quantum description of the interaction of an electron  with a classical magnetic field can be described via  minimal coupling, whereby   the Hamiltonian is modified by adding the vector potential $\bfA$ to the momentum operator. We consider the case of a positively charged particle of mass $m = 1$  in a spatially uniform time varying  magnetic field, 
$$
\mathbf{B}(t) = B(t) \mathbf{e}_3 \,,
$$
where $\mathbf{e}_3$ is a Cartesian unit vector. 

This magnetic field admits the symmetric gauge vector potential 
$$
\mathbf{A} = \frac{B(t)}{2} (-x_2, x_1, 0)\,, 
$$
whence $\bfB=\operatorname{curl}(\mathbf{A})$, and with this choice of gauge, the Hamiltonian $H(t)$ can be expressed as
\bal
    H(t) &= \frac{|\mathbf{p} + \bfA|^2}{2}\nn\\
    &= \frac{p_1^2}{2}+ \frac{p_2^2}{2} + \frac{B(t)^2}{4} (x_1^2 + x_2^2) \nn\\
    &\hspace{0.5cm}+ \frac{B(t)}{2}(x_1p_2 - x_2p_1 ) + {\frac{p_3^2}{2}}\,,
\eal
where the operators $\mathbf{x} = (x_1, x_2, x_3)$ and  $\mathbf{p} = (p_1, p_2, p_3) $ are respectively  the {position operator} and the {momentum operator} related by the identities
\begin{equation}
    [x_i, p_j] = i\hbar \delta_{ij}\, .
\end{equation}

We introduce the operators
\bq
\pi = x_1 p_2 - x_2 p_1\,, \qquad
\kappa = [x_1, p_1]_+ + [x_2, p_2]_+,
\eq
where the anticommutator $[\cdot,\cdot]_+\coloneqq  ab + ba$. We further define the transverse momentum and position by
\bq
p_\perp^2 = p_1^2 + p_2^2, \qquad
x_\perp^2 = x_1^2 + x_2^2,
\eq
which represent, respectively, the squared momentum in the plane orthogonal to the $\mathbf{e}_3$-axis and the squared distance to this axis. With these notations, we obtain
\begin{equation}
    H(t) = \underbrace{\frac{p_\perp^2}{2} + \frac{B(t)^2}{4}x_\perp^2 }_{\textup{Harmonic oscillator}} + \frac{B(t)}{2}\mathbf{\pi} \ + \!\underbrace{\frac{p_3^2}{2}}_{\textup{free motion}} \label{Ham_Charged_par}.
\end{equation}

The Hamiltonian \eqref{Ham_Charged_par} naturally decomposes into transverse and longitudinal contributions with respect to the $\mathbf{e}_3$-axis. The transverse dynamics are governed by the effective Hamiltonian
\[
\frac{p_\perp^2}{2} + \frac{B(t)^2}{4}x_\perp^2 + \frac{B(t)}{2}\mathbf{\pi},
\]
which corresponds to a two-dimensional harmonic oscillator coupled to the angular momentum $\mathbf{\pi}$. In contrast, the longitudinal term $p_3^2/2$ describes a free motion along the $\mathbf{e}_3$-direction and remains completely decoupled from the transverse variables. Since this contribution does not interact with the perpendicular dynamics and does not affect the structure of the dissipative mechanisms under consideration, it can be neglected without loss of generality. Consequently, in the following analysis, we restrict our attention to the reduced Hamiltonian
\bq
H(t) = \frac{p_\perp^2}{2} + \frac{B(t)^2}{4}x_\perp^2 + \frac{B(t)}{2}\mathbf{\pi},
\eq
which captures the essential features of the system.

The Lie algebra $\mathfrak{g}$,  is characterized by the following commutation relations:
\begin{align*}
    [\pi, \kappa] = 0, && [{x}_\perp^2, \pi] = 0,\\
    [{p}_\perp^2 , \pi] = 0, && [{x}_\perp^2, \kappa] = 4i\hbar{x}_\perp^2, \\
    [{p}_\perp^2, \kappa] = -4i\hbar {p}_\perp^2, && [{x}_\perp^2, {p}_\perp^2 ] = 4i\hbar \kappa.
\end{align*}
 
According to  the commutation relations above, the Lie algebra $\mathfrak{g}$ is spanned by the basis $\{{{p}_\perp^2, {x}^2_\perp, \pi,\kappa }\}$. We will impose the conditions $L\in \mathfrak{g}$ and  the conservation of energy as given by  \eqref{dH^q/dtequiv0}. 
For  $L \in \mathfrak{g}$ there exist real coefficients $\beta_1,\beta_2,\beta_3,\beta_4$ such that
\bq
L = \beta_1  {p}_\perp^2 + \beta_2  {x}_\perp^2 + \beta_3 \kappa + \beta_4 \pi.
\eq
According to equation \eqref{eq:coefficents}, and  the fact that the operator $\pi$ commutes with all the elements of the basis of $\mathfrak{g}$, we deduce that $\dot{B} = 0$, i.e. magnetic field is constant in time.

Using  \eqref{eq:to_solve}, we obtain a system of constraints satisfied by the coefficients of the Lindblad operator. More precisely, the coefficients $(\beta_1,\beta_2,\beta_3,\beta_4)$ must satisfy the following system:
\begin{equation}
    \begin{cases}
     \beta_1\beta_2 - \frac{B^2}{2} \beta_1^2 - \beta_3^2= 0\\
    -16\hbar^2\left( -\frac{1}{2}\beta_2^2 + \frac{B^2}{4} \beta_1\beta_2 - \frac{B^2}{4}\beta_3^2\right) = \frac{B(t)\dot B(t)}{2}\\
    \beta_3\left(\frac{1}{2}\beta_2 + \frac{B^2}{4}\beta_1\right) = 0\\
         0 = \dot B
    \end{cases}
    \nonumber
\end{equation}
%
%
Substituting $\dot{B}(t) = 0$ into the second equation above, the system reduces to
\[\begin{cases}
     \beta_1\beta_2 - \frac{B^2}{2} \beta_1^2 - \beta_3^2= 0,\\
     \beta_2^2  - \frac{B^2}{2} \beta_1\beta_2 +\frac{B^2}{2}\beta_3^2 = 0,\\
    \beta_3\left(\beta_2 + \frac{B^2}{2}\beta_1\right) = 0\,.
    \end{cases} 
\]
If \(\beta_2 + \frac{B^2}{2}\beta_1 = 0,\) then  the first equation yields $- B^2 \beta_1^2 = \beta_3^2$, which  can only hold if $B = 0$ and $\beta_3 = 0$, which in turn implies the trivial solution
\[
\beta_1 = \beta_2 = \beta_3 = 0.
\]
On the other hand, if $\beta_3 = 0$, the system reduces to the relation
\bq
\beta_2 = \frac{B^2}{2}\beta_1.
\eq
Substituting this condition into the expression for $L$, we obtain
{ 
\bal
 L &= {\be_1  {p}_\perp^2 + \frac{\be_1 B^2}{2} 
{x}^2_\perp+ \be_4 \pi}
\nn\\
& = \frac{\beta_1}{2} H + (\be_4 - \be_1\frac{B^2}{2}) \pi\,, 
 \eal
for some functions $\beta_1(t)$ and $\beta_4(t)$.
For convenience we set
\bq
c(t) = \beta_4(t) - \frac{\beta_1(t)}{2} B^2(t)\,.
\eq
Thus, the  master equation reads
\begin{align} \label{charged_particle}
i\hbar \frac{d\rho}{dt}
&= [H(t), \rho]
\\
&\quad - i
\Bigg[
\frac{\beta_1(t)^2}{4}\,[H,[H,\rho]]
+ c(t)^2\,[\pi,[\pi,\rho]]
\nonumber\\
&\qquad\quad
+ \frac{\beta_1(t)\,c(t)}{2}
\Big(
[H,[\pi,\rho]] + [\pi,[H,\rho]]
\Big)
\Bigg].
\nn
\end{align}

}

\begin{remark}
    
The elements $ {p}_\perp^2 ,  {x}_\perp^2, \kappa$ generate the $\mathfrak{su}(1, 1)$ algebra. The elements $x_1, x_2, \pi $ form the {Lie algebra of the Euclidean group }$\mathfrak{se}(2)$, By symmetry the same goes for $p_1, p_2, \pi$.
{In the example above, the lie algebra $\mathfrak{g}$ is given by 
\bq
\mathcal{G}= \mathfrak{su}(1, 1) \oplus \operatorname{Span}(\pi).
\eq
}
\end{remark}

By construction, the master equation  \eqref{charged_particle} is thermodynamically consistent, preserving the energy and producing entropy. Since $[H,\pi]=0$, we note that $[L,H]=0$. Consequently, the populations in a common eigenbasis of $H$ and $\pi$ are unchanged during the dynamical evolution, i.e., the  dissipative dynamic does not exchange energy  with the environment.

\subsubsection{ Spin-1/2  in an external magnetic field (Zeeman effect)}

In this example, we consider the interaction between the spin of a localized electron and an external, time-dependent magnetic field. This interaction is described by the Zeeman Hamiltonian
\bq
H(t) = \sum_{i = 1}^3\sigma_i B_i(t)\,,
\eq
where $\mathbf{B}(t) = (B_1(t), B_2(t), B_3(t))$ denotes the applied magnetic field. The operators $\sigma_1$, $\sigma_2$ and $\sigma_3$ are the Pauli matrices, 
\begin{align*}
		\sigma_1 = \frac{\hbar}{2}\begin{pmatrix}0 && 1 \\ 1 && 0\end{pmatrix} && \sigma_2 =\frac{\hbar}{2} \begin{pmatrix} 0 && -i \\ i && 0\end{pmatrix} && \sigma_3 = \frac{\hbar}{2}\begin{pmatrix}1 && 0 \\ 0 && -1\end{pmatrix}, 
	\end{align*}
which generate the Lie algebra $\mathfrak{su}(2)$ and represent the spin degrees of freedom of the electron. This Hamiltonian governs the coherent precession of the spin around the direction of the magnetic field, with a dynamics entirely determined by the geometry and time dependence of $\mathbf{B}(t)$. The Pauli matrices satisfy  the usual commutation relations 
\begin{align*}
    [\sigma_1, \sigma_2] = i\hbar \sigma_3 && [\sigma_2, \sigma_3]= i\hbar \sigma_1 && [\sigma_3, \sigma_1] = i\hbar \sigma_2\,.
\end{align*}


As with the previous examples, we continue under  the assumptions $\bfH_1$ and $\bfH_2$. Let $L\in \mathfrak{g}$ be given by 
\[
L = c_1 \sigma_1 + c_2\sigma_2 + c_3 \sigma_3\,,
\]
where $c_1$ , $c_2$  and $c_3$ are coefficients. Applying equation \eqref{eq:to_solve}, we obtain a system of constraints satisfied by the coefficients of the Lindblad operator

\begin{equation}
    \begin{cases}
         \dot{B_1} &= -\hbar^2[ c_1(c_2B_2 + c_3 B_3) - (c_2^2 + c_3^2)B_1],\\
          \dot{B_2} &= -\hbar^2[ c_2(c_1B_1 + c_3 B_3) - (c_1^2 + c_3^2)B_2],\\
           \dot{B_3} &= -\hbar^2[ c_3(c_1B_1 + c_2 B_2) - (c_1^2 + c_2^2)B_3]\,.
    \end{cases}
\end{equation}

Up to change of coordinates, we can suppose that at the instant $t$ the magnetic field points only in  the direction $\mathbf{e}_1$, i.e.,  $B_2 = B_3 = 0$ and $B_1 > 0$;  hence we obtain the equations
\begin{equation}
    \begin{cases}
         \dot{B_1} &= \hbar^2(c_2^2 + c_3^2) B_1\\
        \dot{B_2} &= -\hbar^2 c_1c_2 B_1\\
        \dot{B_3} &= -\hbar^2  c_1c_3 B_1 
    \end{cases}
\end{equation}
{If $B$ is constant in time (${\p H}/{\p t} = 0$ implies an isolated system)},
we obtain $c_2= c_3 =0$, and hence 
$$L = c_1 \sigma_1$$
Thus, we retrieve the master equation 
\begin{equation}\label{constante_magnetic}
    \Dt{\rho} = \frac{1}{i\hbar} [H, \rho ]  - c_1^2 [\sigma_1, [\sigma_1, \rho ]]\,.
\end{equation}

{On the  other hand, if $B$ varies in  time},
we obtain  
\begin{equation}
\begin{cases}
    c_1^2 &= \frac{\dot{B_2^2} + \dot{B_3^2}}{\hbar^2\dot{B_1}B_1}\\
    c_2^2 &=  \frac{\dot{B_2^2}\dot{B_1} }{\hbar^2(\dot{B_2^2} + \dot{B_3^2})B_1}\\
    c_3^2 &=  \frac{\dot{B_3^2}\dot{B_1} }{\hbar^2(\dot{B_2^2} + \dot{B_3^2})B_1}
\end{cases}\,.
\end{equation}
The condition of the positivity of the square of the  coefficients $c_1, c_2, c_3$, imposes a  condition on the variation of  the magnetic field, viz. 
\begin{equation}
    \dot{B}_1 > 0\,.
\end{equation}
Thus,  the condition for energy conservation is that  the  magnetic field projection along its direction must increase. Consequently, we recover the  master equation at  the instant $t$
\bq\label{time-dependent_magnetic}
    \Dt{\rho} = 
    \frac{1}{i\hbar }[H, \rho] \, + \!\!  \sum_{1 \leq i,j \leq 3} c_ic_j \sigma_i \rho \sigma_j  - \frac{\hbar^2}{4}\!\! \sum_{1\leq i \leq 3} c_i^2 \,  \rho\, .
\eq

In summary, the metriplectic formalism has produced the master equations of \eqref{constante_magnetic} and \eqref{time-dependent_magnetic}. For the equation \eqref{constante_magnetic} $H$ and $L$ are proportional, hence $[L,H] = 0$. Thus, the dissipative dynamics describes pure dephasing in the spin basis aligned with the magnetic field. 

To illustrate this mechanism, we consider a constant magnetic field. We rotate the coordinate system to align the field with the third axis, so that $H=B_3\sigma_3$, where $B_3(t)= \text{constant}$ and $L=c_3\sigma_3$. We then study the evolution of the initial density matrix
\[
\rho(t = 0 ) = \begin{pmatrix}
    1/2 & 1/2\\ 1/2 & 1/2
\end{pmatrix}, 
\]
which represents a pure spin-$\frac12$ state. As $t\rightarrow\infty$ we find that this matrix relaxes to 
\[
\rho(t = \infty) = \begin{pmatrix}
    1/2 & 0 \\ 0  & 1/2 
\end{pmatrix},
\]
 a mixed state with  quantum decoherence. 
 
 Under the dynamics, one can prove that $\frac{\md}{\md t} \Tr(\rho \sigma_3) = 0 $. Hence, the value  of the spin along the third axis is not affected by the interaction with the environment. However, the spin along the first axis begins with value $\Tr(\rho(t = 0) \sigma_1 ) = \hbar/2$, but  evolution leads to the state $\Tr(\rho(t = \infty ) \sigma_1) = 0$, with both values of spin along the first axis being  equally probable. Hence, there is a loss of information because of the dissipative effect   of decoherence, without the loss of energy.

For a time-dependent magnetic field, the Lindblad $L$ generally won't  commute with the Hamiltonian,  $[L,H]\neq 0$. Consequently, the dissipative dynamics in the master equation  \eqref{time-dependent_magnetic}  induces transitions between instantaneous energy levels, in addition to dephasing.

\subsection{Generalized Bloch--Torrey equations}
\label{Bloch-Torrey equation}

The Bloch equations are macroscopic  equations introduced by F.~Bloch in 1946 \cite{bloch1940nuclear} for describing the temporal evolution  of the  nuclear magnetization
\(\mathbf{M} = (M_x, M_y, M_z)\) in the presence of relaxation processes with  longitudinal and transverse relaxation times \(T_1\) and \(T_2\), respectively.
In the presence of an externally applied magnetic field \(\mathbf{H} = (H_x,H_y,H_z)\), the Bloch equations for \(\mathbf{M}\) are 
\begin{equation}
\frac{\partial \mathbf{M}}{\partial t}
=
\gamma\, \mathbf{M} \times \mathbf{H}
-
\begin{pmatrix}
\dfrac{M_x}{T_2} \\[6pt]
\dfrac{M_y}{T_2} \\[6pt]
\dfrac{M_z - M_0}{T_1}
\end{pmatrix}\,,
\end{equation}
where \(\gamma\) is the gyromagnetic ratio, \(M_0\) denotes the equilibrium magnetization, and \(\times\) is usual  the vector cross product. The equilibrium magnetization is assumed to be oriented along the \(z\)-axis.

In the limit \(T_1,T_2 \to \infty\), the relaxation terms vanish and the equations reduce to those for Larmor precession of the magnetization vector in the external magnetic field.  This part is the Hamiltonian/nondissipative part,  while the terms involving $T_1$ and $T_2$ introduced by Bloch  account for dissipative relaxation effects.

A decade later, H.~C.~Torrey extended the Bloch equations by adding diffusion effects associated with the transport of magnetization \cite{torre1956bloch}. This generalization allows for the description of systems with spatial inhomogeneities in the magnetic field, relaxation rates, or initial magnetization. The equations known as the  Bloch--Torrey equations take are 
\bal
\frac{\partial \mathbf{M}}{\partial t}
&=
\gamma\, \mathbf{M} \times \mathbf{H}
-
\begin{pmatrix}
\dfrac{M_x}{T_2} \\[6pt]
\dfrac{M_y}{T_2} \\[6pt]
\dfrac{M_z - M_0}{T_1}
\end{pmatrix}
\nn\\[1.0ex]
&\hspace{1.5cm}+
\nabla \cdot  \big(D\, \nabla (\mathbf{M}-\mathbf{M}_0)\big)\,,
\eal
where \(D\) is the self-diffusion coefficient and \(\mathbf{M}_0=(M_{x_0}, M_{y_0}, M_{z_0})\)  is again the equilibrium magnetization. 

For simplicity  and,  following the assumptions originally adopted by Torrey, we restrict attention to the case of a strong uniform magnetic field applied in the \(z\)-direction, with at most small, spatially varying components in the transverse (equatorial) plane. In this setting, it is natural to take the equilibrium magnetization to be
\[
M_{x_0} = M_{y_0} = 0,
\qquad
M_{z_0} = M_0 ,
\]
where \(M_0\) denotes the equilibrium value of the longitudinal magnetization.
We further assume that the applied magnetic field is uniform and oriented along the \(z\)-axis, namely,
\[
\mathbf{H} = (0,0,H_0)\,.
\]
Under these assumptions, the evolution equation for the longitudinal component \(M_z\) admits the trivial stationary solution 
$M_z = M_0$.  Consequently, the dynamics of \(M_z\) decouples from the transverse components and will not be considered further. In the following, we therefore focus exclusively on the evolution of the transverse magnetization components \(M_x\) and \(M_y\).

{We begin by describing the quantum contribution to the dynamics, focusing on the nondissipative  and dissipative mechanisms encoded in the metriplectic structure.}

\subsubsection{Quantum processes}

To derive the evolution equations for the transverse components of the magnetization, \(M_x\) and \(M_y\) we construct  the Poisson bracket for the nondissipative quantum dynamics, together with the symmetric operator \(\Sigma(\,\cdot\,,\,\cdot\,)\) responsible for dissipation.

We consider a two-level quantum system and choose as a basis for  the space of observables the \(2\times2\) identity matrix \(\mathbb{1}\) together with the three Pauli matrices, 
\[
\sigma_x =
\begin{pmatrix}
0 & 1 \\
1 & 0
\end{pmatrix},
\qquad
\sigma_y =
\begin{pmatrix}
0 & -\mathrm{i} \\
\mathrm{i} & 0
\end{pmatrix},
\qquad
\sigma_z =
\begin{pmatrix}
1 & 0 \\
0 & -1
\end{pmatrix}\,.
\]
Any self-adjoint operator \(\cals\) can then be written as
\bal
\cals &= \frac{1}{2}({\operatorname{Tr}(\cals)}\,\mathbb{1}
+ s_x \sigma_x + s_y \sigma_y + s_z \sigma_z)
\nn\\
& := \frac{1}{2}({\operatorname{Tr}(\cals)}\,\mathbb{1} + \mathbf{s}\cdot \boldsymbol{\si})\,.
\eal

Because the density matrix has unit trace, it admits the representation
\begin{equation}
\rho
=
\frac{1}{2}(\mathbb{1}
+ M_x \sigma_x
+ M_y \sigma_y + M_z \sigma_z)\,,
\end{equation}
whence one can recover  $\mathbf{M}$ from $\rho $ as follows:
\begin{align*}
    M_x = \Tr(\rho\sigma_x) &&  M_y = \Tr(\rho \sigma_y ) &&  M_z = \Tr(\rho \sigma_z) \,,
\end{align*}
where, consistent  with the assumptions of the previous subsection, the longitudinal component is fixed at its equilibrium value. 
The quantum Hamiltonian functional is taken to be
\begin{equation}
\mathcal{H}^q = \operatorname{Tr}(\rho H),
\end{equation}
with Hamiltonian operator
\begin{equation}
{H = \mathbf{H}\cdot\boldsymbol{\sigma}
=   H_0 \,\sigma_z\, ,}
\end{equation}
where $\mathbf{H} = (0,0,H_0)$  is the  uniform magnetic field. In this setting, the dynamics describes a rotation in the \((x,y)\)-plane with no evolution along the \(z\)-direction. The quantum Hamiltonian is given by 
\bq
\mathcal{H}^q = H_0 M_z = \bfH\cdot\bfM\,.
\eq
Alternatively, one could also consider $\mathcal{H}^q =- \bfH\cdot\bfM$, 
which coincides with Zeeman energy. However, this requires a change in sign of  the Poisson bracket, which we choose not to do. The quantum entropy functional is chosen to be
\begin{equation}
\mathcal{S}^q = - \operatorname{Tr}(\rho \ln \rho).
\end{equation}

For the construction of the Poisson bracket, we consider operators of the form
\[
F = \lambda\,\mathbb{1} + \mathbf{f}\cdot\boldsymbol{\sigma},
\qquad
G = \mu\,\mathbb{1} + \mathbf{g}\cdot\boldsymbol{\sigma}.
\]
Using the Pauli matrix algebra, their commutator satisfies
\begin{equation}
[F,G]
=
\,\mathrm{i}\,
(\mathbf{f}\times\mathbf{g})\cdot\boldsymbol{\sigma}.
\end{equation}
Accordingly, the nondissipative  quantum dynamics is governed by
\begin{equation}
\frac{\md \rho}{\md t}
=
-\,\mathrm{i}\gamma\,[H,\rho] 
\end{equation}
where $\ga$ is a measure of the gyromagnetic ratio and  the quantum Poisson bracket is chosen to be 
\begin{equation}
\{\mathcal{A},\mathcal{B}\}_{q}
=
-\,\mathrm{i}\gamma\,
\operatorname{Tr}
\bigl(
\rho\,[\nabla_{\rho}\mathcal{A},\nabla_{\rho}\mathcal{B}]
\bigr),
\label{PBBT}
\end{equation}
which generates the Hamiltonian Larmor precession.

The symmetric operator \(\Sigma\) is constructed as in \eqref{SLIP} using the Bogoliubov inner product of \eqref{BIP}, viz. 
\begin{equation}
\Sigma(F^q,G^q)
=
\sum_{k=1}^{2}
\gamma_k
\left\langle
\big[ L_k , \nabla_{\rho}F^q \big],
\big[ L_k , \nabla_{\rho}G^q \big]
\right\rangle_{\ln \rho}\,,
\nn
\end{equation}
where in the present context the Lindblad operators are chosen as
\[
L_1 = \sigma_x\,,
\qquad
L_2 = \sigma_y\,, 
\]
and the dissipation coefficients are taken to be equal, i.e.,  $
\gamma_1 = \gamma_2 = {1}/({2T_2})$.
Evidently,  we   have 
\bq
\Sigma(\mathcal{H}^q,\mathcal{H}^q) > 0\,.
\eq
From  the Pauli matrix commutation relations, we obtain 
\bq
[\sigma_x,[\sigma_x,\sigma_z]]
= 4\,\sigma_z \quad \mathrm{and}\quad 
[\sigma_y,[\sigma_y,\sigma_z]]
= 4\,\sigma_z\,,
\nn
\eq
which implies
\[
[\sigma_x,[\sigma_x,\sigma_z]]
+
[\sigma_y,[\sigma_y,\sigma_z]]
=
8\,\sigma_z.
\]
Thus,  $\Sigma(\mathcal{S}^q,\mathcal{H}^q)$ becomes as follows:
\begin{align}
\Sigma(\mathcal{S}^q,\mathcal{H}^q)
&=
\operatorname{Tr}\!\left(
\frac{1}{2T_2}
[L_1,[L_1,H]]\,\rho
+
\frac{1}{2T_2}
[L_2,[L_2,H]]\,\rho
\right) \notag\\
&=
\operatorname{Tr}\!\left(
\frac{8 H_0}{T_2}\sigma_z\,\rho
\right)
= \frac{4H_0 M_z}{T_2} 
= \frac{4H_0 M_0}{T_2} .
\end{align}
Note, if  one considers the Zeeman Energy, then 
$$
\Sigma(\mathcal{S}^q,\mathcal{H}^q)
= \frac{-4H_0 M_0}{T_2}\,.
$$
\begin{remark}
If $M_0 = 0$ or $T_2 = \infty$, then $\Sigma(\mathcal{S}^q,\mathcal{H}^q)=0$. According to 
Definition \eqref{definition1}, this implies that the classical environment is unaffected by the quantum subsystem. \label{remark_Envi_uneff}
\end{remark}

We record here, the following interesting lemma:
\begin{lemma}\label{lemma2}
For any quantum observable of the form
\[
\cala^q = \operatorname{Tr}(\rho A),
\quad\text{with }\, 
A \in \operatorname{Span}(\mathbb{1},\sigma_x,\sigma_y),
\]
one has
\[
\Sigma(\cala^q,\mathcal{H}^q)=0\,.
\]
\end{lemma}
This result will play a crucial role in the
analysis of the hybrid quantum--classical dynamics.

\begin{remark} 
\label{R_Quantum_entropy}
Consider the parametrization
\[
\rho(t)=\frac12\big(\mathbb{1}+\mathbf M(t)\cdot\boldsymbol\sigma\big) 
\quad \mathrm{with}\quad
r(t)=|\mathbf M(t)|\,.
\]
The eigenvalues of $\rho$ are $\lambda_\pm=\frac12(1\pm r)$, hence
\[
\mathcal{S}^q(\rho)=-\mathrm{Tr}(\rho\ln\rho)
=-\frac{1+r}{2}\ln\frac{1+r}{2}
-\frac{1-r}{2}\ln\frac{1-r}{2}.
\]
Differentiating with respect to $t$,
\[
\frac{d\mathcal{S}^q}{dt}
=
\frac12\ln\!\left(\frac{1-r}{1+r}\right)\frac{dr}{dt}.
\]
Since $0<r<1$,  
\[
\ln\!\left(\frac{1-r}{1+r}\right)<0,
\]
we obtain
\bq
\frac{d\mathcal{S}^q}{dt}\ge0
\quad\Longleftrightarrow\quad
\frac{dr}{dt}\le0\,.
\label{entM}
\eq
Because $r^2=M_x^2+M_y^2+M_z^2$, \eqref{entM} is equivalent to
\bq
\frac{d}{dt}\big(M_x^2+M_y^2+M_z^2\big)\le0.
\nn
\eq
Physically, an increase of entropy corresponds to a local decrease of the norm of the magnetization vector $\mathbf M$. 
Thus, under the effect of the external magnetic field $\mathbf H$, the magnetization vector $\mathbf M$ relaxes toward its equilibrium configuration while precessing around the magnetic field.
\end{remark}

\subsubsection{Classical processes}

In this subsection, we introduce the classical counterpart of the system, where dissipation is modeled through the metriplectic or  
gradient-flow structure consistent with thermodynamic principles. The magnetic induction is defined as the sum of the 
applied magnetic field and the magnetization,
\begin{equation}
\mathbf{B}= \mu_0 \big( \mathbf{H} + \mathbf{M} \big).
\end{equation}
The classical energy of the magnetic environment is taken in the standard 
form
\begin{equation}
\mathcal{H}^c
=
\int_{\Omega}
\frac{\mathbf{B}^2 }{2\mu_0}\, \mathrm{d}x .
\end{equation}
For the classical entropy, we adopt a Landau-type functional for the magnetization (e.g.,  Eq.~(3.3.3) on  page 84 of  \cite{MorrishMagnetism}  or  \cite{Landau,sab2005magnetocaloric,amaral2010estimating}).  With this choice, the entropy functional is given by
\bal
\mathcal{S}^c[\mathbf{M}]
&=
-\int_{\Omega}\left(
A'(T)|\mathbf{M}|^2/2
+ B'(T)|\mathbf{M}|^4\right.
\nn
\\
&\hspace{2cm}\left.+ C'(T)|\mathbf{M}|^6 \,
\right) \mathrm{d}x\,.
\eal
For simplicity, we assume  $A'(T)=1$ and $B'(T)=C'(T)=0$. 

We construct the symmetric operator $M$ as follows: 
\begin{equation}
M(F^c,G^c)
=
\int_{\Omega}
\nabla \frac{\delta F^c}{\delta \mathbf M}
:
{D}^{\prime}
\nabla \frac{\delta G^c}{\delta \mathbf M}
\, \mathrm{d}x ,
\end{equation}
where $D^{\prime}$ denotes a diffusion coefficient.  The needed functional derivatives are
\begin{equation}
\frac{\delta \mathcal S^c}{\delta \mathbf M} = -\mathbf{M}
\quad\mathrm{and}\quad 
\frac{\delta \mathcal H^c}{\delta \mathbf M}
= \mu_0(\mathbf{M}+\mathbf{H})=\mathbf{B}\,, 
\end{equation}
whence, 
\bq
M(\mathcal S^c,\mathcal H^c) = -\int_\Om\na \mathbf{M}:D^{\prime}\na \mathbf{B}\, \md x \neq 0\,, 
\eq
and 
\bq
M(\mathcal H^c,\mathcal H^c) = \int_{\Om}\na \mathbf{B}:D^{\prime}\na\mathbf{B}\,\md x > 0\, .
\eq
{
\begin{remark}
    
Similarly to Remark~\ref{R_Quantum_entropy}, we obtain
\[
\frac{d\mathcal{S}^c}{dt}\ge0
\quad\Longleftrightarrow\quad
\frac{d}{dt}\int_{\Omega}|\mathbf M|^2\,dx \le 0.
\]
Hence, the increase of the classical entropy is equivalent to the decay of the $L^2$-norm of the magnetization. 
In particular, entropy production is directly associated with the dissipation of magnetic energy 
toward its equilibrium configuration.
\end{remark}
}

Finally, in the next subsection, we combine the quantum and classical contributions to obtain the full hybrid dynamics, highlighting the interplay between the  Hamiltonian and dissipative effects.

\subsubsection{Hybrid process}

Within the metriplectic framework described in Sec.~\ref{Metriplectic formalism for quantum-classical hybrids system},  a hybrid functional  has the form $F=F^q + F^c$,  and its dynamical evolution is described by 
\begin{align}
\dot{F}
=&
\{F^q,\mathcal{H}^q\}_{q}
+
\{F^c,\mathcal{H}^c\}_{c}
+
(F,\mathcal{H};\mathcal{S},\mathcal{H})_{h} \nonumber\\
=&
\{F^q,\mathcal{H}^q\}_{q}
+
\Sigma(F^q,\mathcal{S}^q)\,M(\mathcal{H}^c,\mathcal{H}^c)\nn
\\
&-
\Sigma(F^q,\mathcal{H}^q)\,M(\mathcal{S}^c,\mathcal{H}^c)
+
M(F^c,\mathcal{S}^c)\,\Sigma(\mathcal{H}^q,\mathcal{H}^q)\nn\\
&-
M(F^c,\mathcal{H}^c)\,\Sigma(\mathcal{H}^q,\mathcal{S}^q)\,.
\label{BT_evolution}
\end{align}
For $F^q=\cala = \operatorname{Tr}(\rho A)$ with $ A \in \operatorname{Span}(\mathbb{1},\sigma_1,\sigma_2)$},  using   Lemma \ref{lemma2}, 
the expression of  \eqref{BT_evolution} reduces to
\bal
\dot{F}
=
\{F^q,\mathcal{H}^q\}_{q}
&+
\Sigma(F^q,\mathcal{S}^q)\,M(\mathcal{H}^c,\mathcal{H}^c)\nn\\
&+
M(F^c,\mathcal{S}^c)\,\Sigma(\mathcal{H}^q,\mathcal{H}^q)\nn\\
&-M(F^c,\mathcal{H}^c)\,\Sigma(\mathcal{H}^q,\mathcal{S}^q)\,.
\eal
Using the normalizations
\[
\Sigma(\mathcal{H}^q,\mathcal{H}^q)\equiv 1\quad \mathrm{and}\quad 
M(\mathcal{H}^c,\mathcal{H}^c)\equiv 1\,,
\]
 the hybrid evolution law takes the compact form
\bal
\dot{F}
=
\{F^q,\mathcal{H}^q\}_{q}
&+
\Sigma(F^q,\mathcal{S}^q)
+
M(F^c,\mathcal{S}^c)\nn\\
&-
M(F^c,\mathcal{H}^c)\,\Sigma(\mathcal{H}^q,\mathcal{S}^q)\,.
\label{Hybride_evolution_BT}
\eal

We consider the density matrix
\begin{equation}
\rho
=
\frac{1}{2}(\mathbb{1}
+
\mathbf{M}\cdot\boldsymbol{\sigma})
=
\frac{1}{2}(\mathbb{1}
+
M_x\sigma_x+M_y\sigma_y + M_z \si_z) ,
\end{equation}
corresponding to transverse magnetization in a uniform magnetic field
$\mathbf{H}=(0,0,H_0)$.

The dissipative quantum contribution is governed by the symmetric operator
$\Sigma$. For a quantum observable $F^q=\operatorname{Tr}(\rho A )$ one finds
\begin{align}
\operatorname{Tr}\!\left(\frac{\mathrm{d}\rho}{\mathrm{d}t}\, A \right)
&=
\Sigma(F^q,\mathcal{S}^q)
\\
&=
-\operatorname{Tr}\!\left(
\frac{A}{2T_2}[\sigma_x,[\sigma_x,\rho]]
+
\frac{A}{2T_2}[\sigma_y,[\sigma_y,\rho]]
\right)\,.
\nonumber
\end{align}
The righthand side of the above collapses to
\bal
&-\frac{1}{2T_2}[\sigma_x,[\sigma_x,\rho]]
-
\frac{1}{2T_2}[\sigma_y,[\sigma_y,\rho]]
=
\nn\\
&\hspace{3cm}-\frac{1}{T_2}\left(M_x\sigma_x+M_y\sigma_y\right).
\eal
Since this relation holds for arbitrary $A \in \operatorname{Span}(\mathbb{1},\sigma_1,\sigma_2)$, the transverse components satisfy
\begin{equation}
\partial_t M_x = -\frac{M_x}{T_2}
\quad\mathrm{and}\quad 
\partial_t M_y = -\frac{M_y}{T_2}\,.
\end{equation}

Now, the  classical dissipative dynamics is generated by the symmetric operator $M$ as described above  is given by 
\bal
\partial_t \mathbf{M}
&=
M(\mathbf{M},\mathcal{S}^c) - M(\bfM,\mathcal{H}^c)\Si(\mathcal{H}^q,\mathcal{S}^q)
\nn\\
&=
\nabla\!\cdot\!\bigl(D^{\prime}\,\nabla\mathbf{M}\bigr) + \frac{4H_0 M_0}{T_2}\nabla\!\cdot\!\bigl(D^{\prime}\,\nabla\mathbf{B}\bigr)
\nn\\
& = \nabla\!\cdot\!\bigl(D^{\prime}\,\nabla\mathbf{M}\bigr) + \frac{4H_0 M_0\mu_0}{T_2}\nabla\!\cdot\!\bigl(D^{\prime}\,\nabla\mathbf{M}\bigr)
\nn\\
&= \nabla\cdot(D\nabla\bfM),
\eal
with $D = D^{\prime}(1 +  {4H_0 M_0\mu_0}/{T_2})$, describing the diffusion of magnetization. { We remark that if $M_0 = 0$ or $T_2 = \infty$,  then $D=D^{\prime}$  and this explain how the classical environment is unaffected by the subsystem quantum which is also aligned with the Remark \ref{remark_Envi_uneff}.

Thus we observe that our hybrid metriplectic formalism has combined the nondissipative  Larmor precession with quantum transverse relaxation and classical diffusion to yield a  thermodynamically consistent generalization of  the  Bloch--Torrey equations, i.e.,  
\eqref{Hybride_evolution_BT} yields the following:
\begin{align}
\partial_t M_x
&=
\gamma(\mathbf{M}\times\mathbf{H})_x
- 
\frac{M_x}{T_2}
+
\nabla\!\cdot\!\bigl(D\,\nabla M_x\bigr),
\\
\partial_t M_y
&=
\gamma(\mathbf{M}\times\mathbf{H})_y
-
\frac{M_y}{T_2}
+
\nabla\!\cdot\!\bigl(D\,\nabla M_y\bigr)\,,
\end{align}
where $D = D^{\prime}(1 +  {4H_0 M_0\mu_0}/{T_2})$ mixes diffusion and relaxation in such a way as to conserve the total energy and produce the entropy. 
This  thermodynamically consistent hybrid  system  emerged  naturally from the  metriplectic formalism.

\section{Conclusion}
\label{conclusion}

In this work, we have developed a  metriplectic formulation for quantum--classical hybrid systems, aimed at providing a unified and thermodynamically consistent description of their dynamics. By combining a hybrid Poisson bracket governing the Hamiltonian evolution with a metriplectic $4$-bracket generating dissipative  effects, we constructed a framework that simultaneously ensures conservation of the total energy and non-negative entropy production for the coupled system.

A central feature of the present approach is that the interaction between the quantum subsystem and its classical environment is mediated exclusively through the dissipative part of the dynamics. This choice allows for a transparent interpretation of dissipation in terms of entropy production, and is particularly suited to regimes in which decoherence dominates over coherent coupling. Within this setting, the hybrid entropy naturally decomposes into classical and quantum contributions, providing a clear physical picture of the interplay between microscopic quantum effects and macroscopic classical behavior.

The formalism was illustrated through several applications. In the purely quantum case, we showed that the metriplectic structure leads naturally to the Lindblad master equation, thereby recovering the standard description of open quantum systems from a thermodynamically consistent perspective. In the hybrid setting, we derived the Bloch--Torrey equation as a paradigmatic example of coupled quantum--classical dynamics. This result highlights the ability of the framework to capture both quantum relaxation mechanisms and classical diffusion processes within a single unified description. Moreover, the approach provides an effective Hamiltonian structure compatible with dissipative dynamics, thereby addressing the longstanding issue of reconciling energy conservation with dissipative  evolution.

The present work opens several directions for further investigation. One important extension would be the inclusion of Hamiltonian  coupling between the classical and quantum sectors, allowing for a more general class of hybrid interactions. Another natural direction concerns the application of the framework to more complex systems, such as spatially extended media or systems with nonlinear interactions. Finally, a deeper analysis of the mathematical structure of the hybrid metriplectic brackets, as well as their relation to other approaches to nonequilibrium quantum dynamics, would further clarify the scope and limitations of the present theory.

We hope that the approach developed here provides a useful step toward a systematic and physically consistent description of dissipative quantum--classical systems.

{\section*{Acknowledgements}

A.Z.\ and A.A.\ acknowledges support from the Mohammed VI Polytechnic University.   P.J.M.\ acknowledges support from the DOE Office of Fusion Energy Sciences under DE-FG02-04ER-54742, and from a Forschungspreis from the Alexander von Humboldt Foundation.}





%
%

\appendix
\section{Calculation of $\Sigma(\cala, \mathcal{S}^{q})$} 
\label{Appendix A}
{
\begin{proposition}
Let $\rho$ be a Hermitian, positive definite density operator and $L$ a bounded
operator. Then
\bq
\int_0^1 \rho^{1-\lambda}\,[L,\ln\rho]\,\rho^\lambda\,d\lambda
=
[L,\rho]\,,
\label{main}
\eq
\end{proposition}
\begin{proof}
\textbf{Method 1:}
Since $\rho$ is Hermitian and positive definite, the spectral theorem ensures
that there exists an orthonormal basis $\{ |i\rangle \}$ and strictly positive
eigenvalues $\{ r_i \}$ such that $\rho = \sum_i r_i\,|i\rangle\langle i|$\,. Consequently, using 
\[
\ln\rho = \sum_i \ln(r_i)\,|i\rangle\langle i|
\qquad\mathrm{and}\qquad
\rho^\lambda = \sum_i r_i^\lambda\,|i\rangle\langle i|
\,,
\]
we can compute the commutator
\[
[L,\ln\rho] = L\ln\rho - \ln\rho\,L 
\]
by inserting the spectral decompositions, yielding
\[
[L,\ln\rho]
=
\sum_{i,j} \bigl(\ln r_j - \ln r_i\bigr)
\,|i\rangle\langle i|L|j\rangle\langle j|\,.
\]
Using this, the integrand of \eqref{main} becomes
\[
\rho^{1-\lambda}[L,\ln\rho]\rho^\lambda
=
\sum_{i,j}
r_i^{1-\lambda} r_j^\lambda
(\ln r_j - \ln r_i)
\,|i\rangle\langle i|L|j\rangle\langle j|.
\]
Thus, for  fixed indices $i,j$,
\bal
& \int_0^1 r_i^{1-\lambda} r_j^\lambda (\ln r_j - \ln r_i)\,d\lambda\\
&\hspace{2cm} =
r_i(\ln r_j - \ln r_i)
\int_0^1 \left(\frac{r_j}{r_i}\right)^\lambda d\lambda\,.
\nn
\eal
For any $r_j/r_i>0$ and  $r_j/r_i\neq 1$,
we have 
\[
\int_0^1 \left(\frac{r_j}{r_i}\right)^\lambda d\lambda
=
\frac{\frac{r_j}{r_i}-1}{\ln r_j - \ln r_i}
\]
and, therefore, 
\[
\int_0^1 \rho^{1-\lambda}[L,\ln\rho]\rho^\lambda\,d\lambda
=
\sum_{i,j} (r_j - r_i)
\,|i\rangle\langle i|L|j\rangle\langle j|\,.
\]
Directly, the commutator $[L,\rho]=L\rho-\rho L$ is 
\[
[L,\rho]
=
\sum_{i,j} (r_j - r_i)
\,|i\rangle\langle i|L|j\rangle\langle j|\,.
\]
Comparison of  the two expressions proves the desired identity.

\medskip

\textbf{Method 2:}
Let 
\[ g: \lambda \mapsto \rho^{1 - \lambda} L \rho ^\lambda \]
and consider its first derivative 
\begin{align*}
    g'(\lambda) &= - \rho^{1 - \lambda} \ln(\rho) L \rho^\lambda + \rho^{1 -  \lambda} L \ln(\rho) \rho^\lambda\\
    &= \rho^{1 - \lambda} [L, \ln(\rho)]\rho^\lambda \,.
\end{align*}
Thus upon integrating
\[ \int_0^1  \rho^\lambda [L, \ln(\rho) ] \rho^{1 - \lambda } = \int_0^1 g'(\lambda)\mathrm{d}\lambda = g(1) - g(0) \,,  \]
which yields the desired equality 
\[\int_0^1 \rho^{1 - \lambda} [L, \ln(\rho) ] \rho^\lambda \mathrm{d}\lambda = [L, \rho]\]
\end{proof}
}
\begin{lemma}
The following are  identities:
    \begin{equation}
        \big[L, [L, \rho ]\big] = \frac{d^2}{dt^2} e^{Lt} \rho e^{-Lt} \Bigg|_{t = 0}
    \end{equation}
    \begin{equation}
        \int_0^1 \rho^\lambda A \rho^{1 - \lambda} d \lambda = \frac{d}{dt} e^{\ln(\rho) + tA}\Bigg|_{t = 0}
    \end{equation}
    \begin{equation}
        \langle A, B\rangle_{ \ln(\rho)} = \frac{d^2}{dt ds} e^{\ln(\rho) + s A+ t B}\Bigg|_{t = s = 0}
    \end{equation}
\end{lemma}
\begin{proof}
These all follow by  expanding the exponential  to second order.
\end{proof}

\section{The increase of the Tsallis entropy}
\label{sec:Tsalli}

\begin{proposition}
    If all the Lindbald operators are Hermtian, $L_i^\dagger = L_i$,  then the Tsallis entropy $\mathcal{S}_q$  defined by \eqref{eq:Tsalli} is increased by the  Lindbald equation, i.e.
    \[ \Dt{\mathcal{S}_q}\geq 0\,.\]
\end{proposition}
\begin{proof}
    Without loss of generality  we set  $L= L_1$ and we  assume $L_2 = L_3 = \cdots = 0$. For  convenience we work in a basis where  $\rho$ is diagonal.  Let $r_i$ be the diagonal elements of $\rho$ and  $L_{ij}$ be the coefficients of $L$ in this basis (we can think of $(L_{ij})$ as infinite matrix).  Differentiation of $\mathcal{S}_q$ gives
    \begin{align*}
        \Dt{\mathcal{S}_q} &= \Tr\left( \frac{q}{1 - q}\rho^{q - 1}\Dt{\rho}\right)\\
        &= \frac{q}{q - 1} \Tr\left(\frac{1}{i\hbar} [H, \rho] \rho^{q - 1} \right.\\
        &\left. + c\left(L^\dagger \rho L -  \frac{1}{2} \{ L^\dagger L, \rho\})\rho^{q - 1}\right) \right)  \\
        &= \frac{q}{q - 1} \underbrace{\Tr\left( \frac{1}{i\hbar} [\rho, \rho^{q -1}]H\right) }_{= 0}\\
        &+c \frac{q}{q - 1}\Tr\left(L^\dagger \rho L \rho^{q - 1} - L^\dagger L \rho ^q\right) 
         \\
        &= \frac{q}{q - 1} c\sum_{i,j} (r_i r_j^ {q- 1} - r_j^q)|L_{ij}|^2\\
        &= c\frac{q}{q - 1} \sum_{i< j} (r_i - r_j)(r_i^{q - 1} - r_j^{q -1})|L_{ij}|^2
    \end{align*}
Because  the function $r \mapsto r^{q}$ is convex,  $r\mapsto r^{q - 1}$ is increasing, which implies  the desired result, i.e., the increase of entropy $\mathcal{S}_q$.
\end{proof}

\section{Proof of Lemma \ref{lemma2}}

To prove that $\Sigma(\cala ,H^q)=0$ for $\cala=\mathrm{Tr}(\rho A)$ with
$A\in \mathrm{Span}(I_2,\sigma_x,\sigma_y)$, let
\[
\rho=\tfrac12 (\mathbb{1}+M_x\sigma_x+M_y\sigma_y),
\qquad
A=a_0 \mathbb{1}+a_x\sigma_x+a_y\sigma_y .
\]

\paragraph{Step 1: Commutators}
\[
[\sigma_x,A]=a_y[\sigma_x,\sigma_y]=2i\,a_y\sigma_z,
\qquad
[\sigma_x,\sigma_z]=-2i\,\sigma_y .
\]

\paragraph{Step 2: Inner product}
\[
\langle[\sigma_x,A],[\sigma_x,\sigma_z]\rangle_{\ln\rho}
=
4a_y
\int_0^1 d\lambda\;
\mathrm{Tr}\!\left(
\rho^\lambda\sigma_z\rho^{1-\lambda}\sigma_y
\right).
\]

\paragraph{Step 3: Vanishing of the trace}
Since the Bloch vector of $\rho$ lies in the $(\sigma_x,\sigma_y)$-plane,
any power of $\rho$ has the form
\[
\rho^\lambda=\alpha_\lambda I_2
+\beta_\lambda\bigl(M_x\sigma_x+M_y\sigma_y\bigr).
\]
Each term in
$\mathrm{Tr}(\rho^\lambda\sigma_z\rho^{1-\lambda}\sigma_y)$
contains exactly one $\sigma_z$ and one $\sigma_y$. By the Pauli trace
identities,
\[
\mathrm{Tr}(\rho^\lambda\sigma_z\rho^{1-\lambda}\sigma_y)=0\,, 
\qquad \forall\,\lambda \,.
\]
To see this, let
\[
\rho=\tfrac12 (\mathbb{1}+M_x\sigma_x+M_y\sigma_y)
=\tfrac12( I_2+\bfM\cdot\boldsymbol{\si}).
\]
Note
\[
\begin{aligned}
M^2
&=M_x^2\sigma_x^2
+M_xM_y\sigma_x\sigma_y
+M_xM_y\sigma_y\sigma_x
+M_y^2\sigma_y^2 \\
&=(M_x^2+M_y^2)I_2
=:r^2 I_2\,,
\end{aligned}
\]
using
\[
\sigma_x\sigma_y=i\sigma_z=-\sigma_y\sigma_x,
\qquad
\sigma_x^2=\sigma_y^2=I_2 \,.
\]
The eigenvalues of $\rho$ are
\[
\lambda_\pm= {(1\pm r)}/{2},
\]
with spectral projectors
\[
P_\pm=\tfrac12\!\left(I_2\pm\frac{M}{r}\right).
\]
For any analytic function $f$,
\[
f(\rho)=f(\lambda_+)P_+ + f(\lambda_-)P_-,
\qquad
\rho^\lambda=\lambda_+^\lambda P_+ + \lambda_-^\lambda P_-.
\]
Substituting the projectors,
\[
\rho^\lambda
=\tfrac12(\lambda_+^\lambda+\lambda_-^\lambda)I_2
+\tfrac12(\lambda_+^\lambda-\lambda_-^\lambda)\frac{M}{r}.
\]
Therefore,
\[
\rho^\lambda
=a_\lambda I_2
+b_\lambda(M_x\sigma_x+M_y\sigma_y),
\]
with
\[
a_\lambda=\tfrac12(\lambda_+^\lambda+\lambda_-^\lambda),
\qquad
b_\lambda=\frac{\lambda_+^\lambda-\lambda_-^\lambda}{2r}.
\]
Since $\rho$ contains no $\sigma_z$ term and $M^2\propto I_2$,
no power of $\rho$ can generate a $\sigma_z$ component.

\paragraph{Final Step}:
\[
\langle[\sigma_x,A],[\sigma_x,\sigma_z]\rangle_{\ln\rho}=0
\]
The same argument with $\sigma_x\leftrightarrow\sigma_y$ yields
\[
\langle[\sigma_y,A],[\sigma_y,\sigma_z]\rangle_{\ln\rho}=0
\]
Therefore,
\begin{equation}
\Sigma(\cala,\mathcal{H}^q)
=
\sum_{k=x,y}\frac{1}{2T_2}
\langle [\si_k,A],[\si_k,H]\rangle_{\ln\rho} = 0\,.
\end{equation}

\bibliographystyle{unsrt}




%




\end{document}